\documentclass{article}

\usepackage{arxiv}

\usepackage[utf8]{inputenc} 
\usepackage[T1]{fontenc}    
\usepackage{hyperref}       
\usepackage{url}            
\usepackage{booktabs}       
\usepackage{amsmath}
\usepackage{amsfonts}
\usepackage{amssymb}
\usepackage{graphicx}
\usepackage{subfigure}
\usepackage{doi}
\usepackage{textgreek}
\usepackage{authblk}
\usepackage[square,numbers]{natbib}

\title{Visualizing particle networks in granular media by in situ X-ray computed tomography}

\date{December 23, 2021}

\author[1,*]{Matthias Ruf}
\author[1,2]{Kianoosh Taghizadeh}
\author[1,3]{Holger Steeb}
\affil[1]{Institute of Applied Mechanics~(CE), University of Stuttgart, Stuttgart, Germany}
\affil[2]{Multiscale Mechanics~(MSM), University of Twente, Enschede, The Netherlands}
\affil[3]{SC SimTech, University of Stuttgart, Stuttgart, Germany}
\affil[*]{Corresponding author: Matthias Ruf, \url{matthias.ruf@mechbau.uni-stuttgart.de}}
\date{December 23, 2021}

\renewcommand{\headeright}{A preprint - December 23, 2021}
\renewcommand{\undertitle}{A preprint}
\renewcommand{\shorttitle}{}

\hypersetup{
pdftitle={Visualizing particle networks in granular media by in situ X-ray computed tomography},
pdfsubject={cond-mat.mtrl-sci, physics.geo-ph, physics.ins-det},
pdfauthor={Matthias Ruf, Kianoosh Taghizadeh, Holger Steeb},
pdfkeywords={granular media, wave propagation, X-ray computed tomography},
}

\begin{document}
\maketitle

\begin{abstract}
In the present contribution, cylindrical samples consisting of monodisperse soft (rubber) and stiff (glass) particles are pre-stressed under uniaxial compression. Acoustic P-waves at ultrasound frequencies are superimposed into prepared samples with different soft-stiff volume fraction. Earlier investigations showed the importance of particles networks, i.e. force chains, in controlling the effective mechanical properties of particulate systems, e.g. elastic moduli. Measured P-wave modulus showed a significant decline while more soft particles are added, i.e. higher rubber fractions, due to a change in microstructure. However, for small contents of soft particles, it could be observed that the P-wave modulus is increasing. For the understanding of such kinds of effects, detailed insight into the microstructure of the system is required. To gain this information and link it to the effective properties, we make here use of high-resolution micro X-ray Computed Tomography ({\textmu}XRCT) imaging and combine it with the classical stiffness characterization. Both performed in situ meaning inside the laboratory-based XRCT scanner. With {\textmu}XRCT imaging, the granular microstructure can be visualized in 3d and characterized subsequently.  By post-processing of the {\textmu}XRCT data, the individual grains of the particulate systems could be uniquely identified. Finally, the contact network of the packings with low and high rubber contents which connects the center of particles was established to demonstrate the network transition from stiff- to soft-dominated regimes. This has allowed for unprecedented observations and a renewed understanding of particulate systems. It has been demonstrated that {\textmu}XRCT scans of particles packings can be analyzed and compared in 3d to gain extensive information on the scale of the single particles. Here, the in situ setup and workflow from the start of acquiring images in situ till the post-processing of the image data is explained in detail and demonstrated by selected results.
\end{abstract}

\keywords{granular media \and wave propagation \and X-ray computed tomography}

\section{Introduction}
Dense granular media are of widespread importance in a number of applications, ranging across time and
length scales from geophysical earthquakes in the San Andreas fault zone, to plastic powder sintering
in the 3d printer on your desk. Granular mixtures are of interest for a large number of fields, materials, and applications, including mineral processing, environmental engineering, geomechanics and geophysics, and have received a lot of attention in the last decades. A specific example in geotechnical engineering is the increasing incorporation of recycled materials (e.g.~shredded or granulated rubber, crushed glass) often used into conventional designs and soil improvement projects \cite{bosscher1997design,garga2000tire,lee1999shredded}. Moreover, sophisticated mixtures of asphalt and concrete are widely used to construct roads \cite{heimdahl1999elastic,hinisliouglu2004use,siddique2004properties,xiao2007rutting,zornberg2004behaviour}.

The mechanical behavior of granular materials is highly nonlinear and involves plastic deformations also for very small strain due to rearrangements of particles \cite{kruyt2010micromechanical,magnanimo2008characterizing,makse2004granular}. On the other hand, the concept of an initial purely elastic regime at small strains for granular assemblies is an issue still under debate in the soil mechanics community   \cite{burland1989ninth,clayton2011stiffness,walton1987effective}. On the other hand, approaches that neglect the effect of elastic stored energy are also questionable, i.e.~all the work done by the internal forces is dissipated. Features visible in experiments, like the propagation of acoustic waves, can hardly be described without considering an elastic regime \cite{taghizadeh2015understanding,taghizadeh2017applied}.  
The small-strain stiffness provides useful soil information, which is relevant to a wide range of engineering applications including the design of foundations subjected to dynamic loading, process monitoring, liquefaction assessment, and soil improvement control. Within the small strain region, the geomaterial exhibits linear-elastic
behaviour and the moduli are independent of strain amplitude \cite{chang1995estimates,jovivcic1997stiffness,kruyt2010micromechanical}. The elastic range depends on confining pressure for non-plastic soils. 

Probing a granular media with (ultra) sound waves gives useful information on the state, the structure and the mechanical properties of the bulk media as there is a one to one relation between a wave speed and small-strain stiffness of packings. 
Estimation of stiffness has traditionally been made in a triaxial apparatus using precise
displacement transducers or resonant column devices \cite{atkinson1991experimental,santamarina1999small}. However, these methods have a disadvantage of destructing samples, whereas ultrasonic measurements (by propagation of an elastic wave) are widely accepted
for their rapid, non-destructive, and low-cost evaluation methods. 
Earlier studies using a wave propagation technique {} have shown that the dissipative, elastic and lightweight properties of materials
(like soils, asphalt, etc.) can be enhanced by deliberately adding dissipative, soft, light inclusions of various types and compositions. Therefore, such a tailored system with improved material properties leads to new application designs.  

There exists a diverse array of experimental imaging techniques which can be exploited in order to investigate particulate systems. One of them is attenuation-based micro X-ray Computed Tomography ({\textmu}XRCT) \cite{Buzug2008,Carmignato2018,Kak1987,Russo2018,Stock2008}. {\textmu}XRCT imaging is in general a non-destructive imaging technique that offers the possibility to visualize the internal structure of objects. In contrast to other microscopy imaging methods like optical microscopy, it provides a 3-dimensional (3d) representation of the investigated sample. The method is based on the physical effect of X-ray attenuation which depends among other things on the atomic number and consequently on the chemical composition in each material point. The final, so called reconstructed 3d volume resulting from an XRCT scan is typically represented as a stack of 2-dimensional (2d) gray values images. Through subsequent post processing of the 3d raw image data, different material phases can be identified and subsequently segmented \cite{Iassonov2009,Russ2016,Schlueter2014,Tuller2015}. Based on the segmented 3d volume, quantifications on different length scales can be performed. One example to be mentioned on the macroscopic level is the determination of the volume fractions of the individual phases as well as their distribution. But it is also possible to obtain detailed quantities such as the coordination number of individual particles in the case of granular materials. Consequently, experimental results and observed phenomena obtained by classical characterization methods can be better and more comprehensively understood by help of {\textmu}XRCT. If {\textmu}XRCT imaging is combined with classical characterization methods and performed at the same location we talk about in situ {\textmu}XRCT imaging \cite{Buffiere2010}. Motivated by the example of a discovered not fully understood effect shown by tailored granular media \cite{taghizadeh2021elastic}, a workflow for in situ investigation is demonstrated in this contribution.

In the following, we start with a short overview of image-based characterization using X-ray computed tomography. The combination (in situ) of this method with the traditional ultrasonic through-transmission measurement technique is the focus of Section~\ref{sec:In-situ_experimental_testing}. We apply this to investigate monodisperse particulate systems consisting of weak (rubber) and stiff (glass) particles of different volume fractions. We show a possible realization including the required hardware as well as the overall workflow. Based on this, some selected results of the investigation are provided in Section~\ref{sec:results} to give a motivation of what is in general possible. The acquired experimental data for one selected mixture is open access published for demonstrative purposes. The proposed in situ workflow is discussed in Section~\ref{sec:discussion}. A summary of the presented work is given in Section~\ref{sec:summary}.

\section{Image-based characterization using X-ray computed tomography}
\label{sec:image_characterization}
\begin{figure}[ht]
	\centering
	\includegraphics[width=0.8\textwidth]{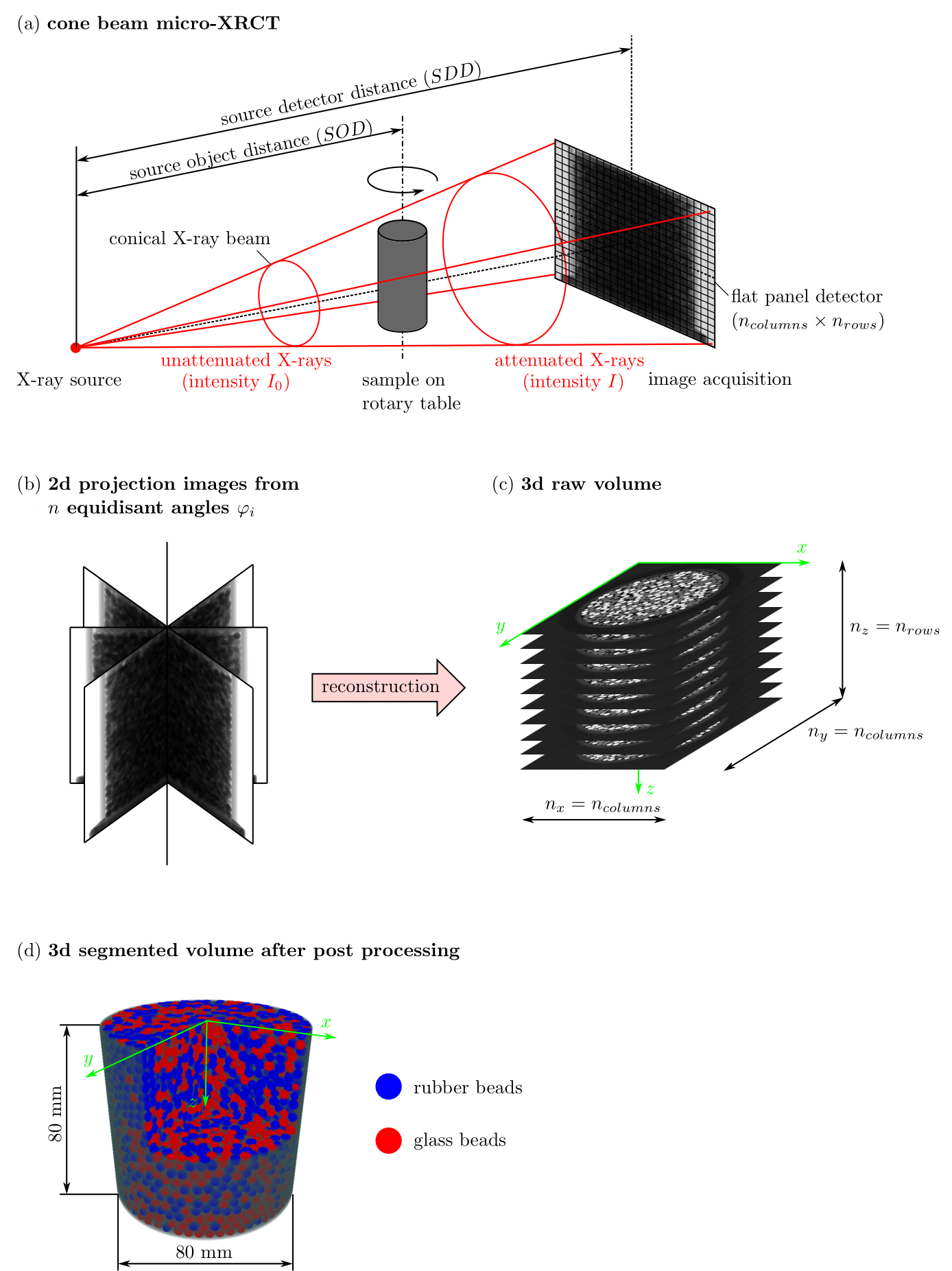}
	\caption{Illustration of the principle of attenuation-based X-ray computed tomography (XRCT) imaging based on the scan of a cylindrical sample of monodisperse stiff (glass) and soft (rubber) particles.}
	\label{fig:xrct_basisc}
\end{figure}
The capability which micro X-ray Computed Tomography ({\textmu}XRCT) as a non-destructive imaging technique offers are meanwhile well known in many research domains and it is becoming more and more a routine microscopy technique \cite{Cnudde2013,Kotwaliwale2011,Landis2010,Maire2013,Mizutani2012,Schoeman2016,Stock2008,Stock2008a}. The method is based on the revolutionary discovery of X-ray radiation by Wilhelm Röntgen in 1895 \cite{Russo2018}. X-rays are high-energy electromagnetic waves with a wavelength in the range of \mbox{0.01 nm} to \mbox{10 nm}. They have a long absorption length and interact with matter through different mechanisms (photoelectric effect, Compton scattering, Rayleigh/Thomson scattering, and pair production) \cite{Carmignato2018}. Depending on the material penetrated, X-rays are attenuated to different degrees. Assuming a monochromatic X-ray beam, the Beer-Lambert law formulates mathematically the transmitted intensity $I$ of an X-ray beam penetrating an object along the straight ray path from 0 to $\tilde{s}$ by
\begin{equation}
	I(\tilde{s}) = I_0 \mathrm{e}^{-\int_0^{\tilde{s}} \mu(s) \mathrm{d}s}
	\label{eq:Beer_Lambert_law}
\end{equation}
where $I_0$ is the initial intensity of the X-ray beam and $\mu(s)$ are the unknown local linear attenuation coefficients along the ray path \cite{Buzug2008,Carmignato2018,Stock2008}. If an object is irradiated from one side and the attenuated X-ray beam is measured on the opposite side by an X-ray detector, a so-called radiogram or projection image results. The distribution of the X-ray intensity~$I$ is typically represented as a gray value image as displayed in Figure~\ref{fig:xrct_basisc}(a) showing one projection image of a cylindrical sample containing monodisperse soft (rubber) and stiff (glass) particles. According to Equation~\eqref{eq:Beer_Lambert_law} each pixel of such an image represents the intensity $I$ of all attenuated X-ray photons which are captured by the specific sensor pixel during the adjusted exposure time. The darker the gray value the less the intensity $I$ and vice versa in the radiogram. Since each pixel contains just the integral information for one specific X-ray path, it is not possible to back-calculate the unknown linear attenuation values in the material points from one radiogram. However, if a sufficiently large set of projections is taken from different directions, cf. Figure~\ref{fig:xrct_basisc}(b), enough information is available to calculate the attenuation coefficient for each material point, cf. Figure~\ref{fig:xrct_basisc}(c). The 3d distribution of the calculated attenuation coefficient then represents the inner structure of the investigated object. It is to mention, due to different reasons, that the absolute values of the 3d images are mostly only correlating with the theoretical attenuation coefficients and are not identical.

The general method for the determination of the 3d structure based on numerous projection images is called reconstruction \cite{Buzug2008,Kak1987}. To capture projections from different directions, in {\textmu}XRCT imaging typically the sample is rotated, cf. Figure~\ref{fig:xrct_basisc}(a) and (b). The number $n$ of required equidistant projection angles $\varphi_i$ over one turn of the sample correlates with the number of detector pixel columns. The size/discretization of the 3d volume is given by the resolution of the detector. The base area (x- and y-direction) corresponds with the number of detector pixels in the horizontal direction and the maximum height (z-direction) with the number of vertical detector pixels, cf. Figure~\ref{fig:xrct_basisc}(c). Corresponding to the term pixel in 2d, in 3d the term voxel is used \cite{Russ2016}. Assuming an identical voxel edge length, the edge length of one voxel is given by the ratio of the detector pixel size and the applied geometric magnification $M_{geo.}$. The adjustment of the magnification in a cone beam system is performed by geometric magnification. If we assume the source-detector-distance ($SDD$) is fixed, the geometric magnification $M_{geo.}$ is adjusted by varying the source-object-distance ($SOD$) and is given by the ratio $M_{geo.} = SDD / SOD$. The maximum theoretically achievable spatial resolution correlates with the underlying focal spot size of the used X-ray source for the adjusted power setting \cite{Stock2008}. A micro-focus X-ray source typically has a focal spot size in the micrometer range and results in a theoretical system resolution which is also in the micrometer range.

For the reconstruction itself, it is distinguished between two groups of reconstruction techniques. On the one side, there are analytical reconstruction techniques and on the other side algebraic ones, e.g. \cite{Buzug2008,Kak1987}. While the first group of methods tries to solve the inverse problem analytically, the second group treats it as an optimization problem. Both have their respective advantages and disadvantages. The shape of the X-ray beam has a significant influence on the reconstruction, in particular for the analytical techniques. The ideal case of a parallel X-ray beam is only present in beamlines of synchrotron radiation facilities \cite{Rahimabadi2020} and not given in laboratory XRCT systems. To account for the conical beam shape in laboratory XRCT systems, the so-called FDK algorithm is mostly applied. The FDK approach is a practical analytical cone-beam algorithm which goes back to Feldkamp L.A., Davis L.C., and Kress J.W. \cite{Feldkamp1984}. Independent of the used reconstruction techniques, the final raw data set consists of a stack of slices called tomograms, cf. Figure~\ref{fig:xrct_basisc}(c). The voxel values, typically represented as gray values, represent their influence on X-ray attenuation. The brighter the are the higher their influence on the X-ray attenuation corresponding with the material density. The bright gray voxels in Figure~\ref{fig:xrct_basisc}(c) correspond to the stiff (glass) particles and the dark gray ones to the (soft) rubber particles and the nearly black voxels reproduce the air phase. The reconstructed data set contains a certain amount of noise due to the procedure as well as potentially other hardly avoidable artifacts \cite{Carmignato2018,Russo2018,Stock2008}. 

By subsequent image processing, based on the 3d raw volume, different kinds of information can be extracted from the data set. For this purpose, a so-called segmentation of the different phases is usually carried out. Segmentation is the partitioning of a gray-scale image into disjoint regions that are homogeneous with respect to some characteristics \cite{Iassonov2009,Russo2018,Schlueter2014,Tuller2015}.

In the simplest case, this can be achieved by defining a threshold value, since the individual phases differ in terms of the average gray value. To determine the related threshold values, a histogram of the underlying gray value distribution of the raw data set is usually employed. The most well-known approach based on this scheme is the ``Otsu method'' \cite{Otsu1979}. However, due to the intrinsic present noise in the raw data set, incorrect assignment of different individual voxels occurs. To account for this, various filtering techniques are used \cite{Iassonov2009,Schlueter2014,Tuller2015}. To mention is that in the last years more and more image segmentation using machine learning techniques has been robustly studied as a new approach due to its well-known benefits \cite{Karimpouli2017,Kodym2018}. Which phases have to be separated depends on the scientific question. Exemplary, in Figure~\ref{fig:xrct_basisc}(d) the separation of the soft (rubber) and stiff (glass) particles, as well as the pore space (transparent), is showed.

By repeating the scanning process as well as the image processing, the evolution of physical processes (e.g. deformation, crack initiation, and growth) over time can be observed. The final result is one stack of tomograms (3d image) for each time step. What kind of processes can be observed depends on the required acquisition time for one complete scan. While acquisition times of less than one second for one complete scan are achievable in beamlines of synchrotron radiation facilities \cite{Buffiere2010,Hasan2020,Rahimabadi2020}, the acquisition time in a standard lab-based XRCT system is between some minutes up to several hours, correlating with the required scan quality. This is one limitation of physical processes which can be studied. However, processes which can be stopped in an equilibrium state are also open for investigation in a laboratory system. For time-resolved scans, the term 4-dimensional (4d) scanning (space + time) is frequently used.

\section{In situ experimental testing}
\label{sec:In-situ_experimental_testing}
The combination of XRCT imaging with classical mechanical characterization methods is an excellent technique to better understand the behavior of materials as well as physical processes in general, cf. \cite{Buffiere2010, Singh2014}. In this context often the term ``in situ X-ray computed tomography'' is used. ``In situ'' is a Latin phrase and translates literally to ``on-site" or ``on place" and is the antonym of ``ex situ''.
In the experimental mechanics' context, it describes the way the measurement is taken, highlighting that the measurements are acquired in the same place the phenomenon is occurring without removing/installing the sample each time. Thus, in situ XRCT means that the mechanical characterization is performed inside the scanner. It offers the possibility to enrich the information on the macroscopic level, e.g. resulting from acoustic measurements as demonstrated within this contribution, with information on the microscopic level resulting from 3d imaging. In the following, we provide an overview of the applied workflow and used hardware to investigate particulate systems with acoustic wave measurements combined with in situ XRCT imaging.

\subsection{Stiffness determination based on wave propagation measurement}
\label{sec:Stiffness_measurement}
Mechanical (sound) waves are disturbances that propagate through space and time in a medium in which deformation leads to elastic restoring forces. 
This produces a transfer of momentum and energy from one material point to another, usually involving little or no associated mass transport if the amplitude is small enough. The P-wave, or primary wave, is the fastest and the first wave detected by seismographs. They are able to move through both solid rock as well as through liquids. P-waves are compressional or longitudinal waves that oscillate the ground back and forth along the direction of wave travel, in much the same way that sound waves (which are also compressional) move air back and forth as the waves travel from the sound source to a sound receiver. In a longitudinal wave, the particle displacement is parallel to the direction of wave propagation \cite{misra2019longitudinal,nesterenko1984propagation,shrivastava2017effect,taghizadeh2021stochastic,mouraille2008sound}.
\begin{figure}[ht]
	\centering
	\includegraphics[width=0.7\textwidth]{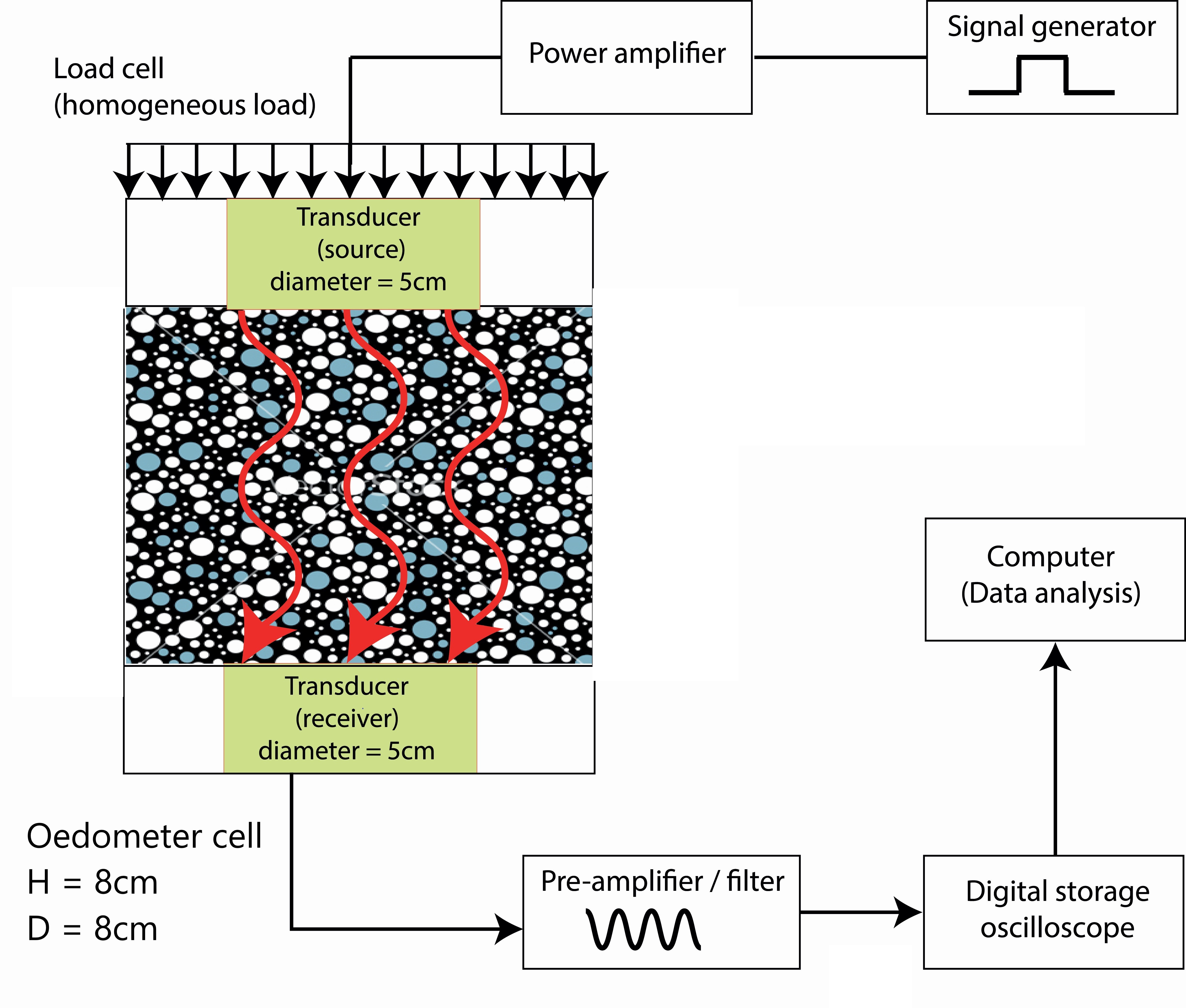}
	\caption{Schematic view of wave propagation measurement setup  \cite{taghizadeh2021elastic}.}
	\label{fig:wave_setup}
\end{figure}
 
Quantitative ultrasonic measurement has been widely-used by different disciplines ((geo-)physics, soil and geomechanics, materials sciences, mechanical and civil engineering) to describe the small strain stiffness behavior of a particulate system, as can be found in literature \cite{jia2001sound,jia1999ultrasound,van2016sound,somfai2005elastic}. 
Velocity testing by through transmission ultrasound methods has gained popularity due to its relative ease of obtaining the modulus of a sample.
Figure \ref{fig:wave_setup} shows a schematic view of the used low-frequency ultrasound setup integrated into an oedometer cell including its electrical pieces, cf. Figure~\ref{fig:experimental_setup} and Figure~\ref{fig:experimental_setup_oedometer_cell}. For our highly attenuating particulate system, it consists of a pair of 100 kHz P-wave broadband piezoelectric transducers (Olympus-Panametrics Videoscan V1011), an ultrasonic square wave pulser/receiver unit (Olympus-Panametrics 5077PR) and a digital oscilloscope (PicoScope 5444B). Piezoelectric transducers are used to determine the small-strain compression stiffness, $M$, of the granular system by determining the velocity of mechanical waves (``speed of sound'') through the tested samples. 
The transducers are generally used in pairs when one transducer operates as a transmitter and the other as a receiver. The transmitting transducer is generally embedded at one end of the particulate soil sample and the receiving is, aligned with the transmitter, embedded at the other end. This kind of pair-wise transducer arrangement is also named ultrasonic through transmission measurement technique in the literature.

This allows to probe the stiffness of the soil or particulate sample along a given stress path. The transmitting transducer transform the input electric signal to an acoustic wave and sends it through the medium. The receiving transducer receives the propagated mechanical sound wave and transforms it back to an electronical signal which could be acquired by the digitizer, compare \cite{lee2005bender,sawangsuriya2012wave} for technical details on the transducers.  Transducers are often incorporated in geomechanical and geophysical laboratory testing equipment such as in triaxial or oedometers cells.

The P-wave velocity ($V_P$) can be calculated from the travel time ($t_P$), given the height ($H$) of the sample in the actual configuration, $V_P = H/t_P$. Knowing the elastic wave velocity ($V_P$) and knowing the total mass density ($\rho_B$) of the particulate system, the longitudinal P-wave stiffness of the sample ($M$) is determined, $M=\rho_B V_P^2$.
In our investigations, the sample consists of rubber and glass particles $\rho_B = \nu \rho_r + (1-\nu)\rho_g$, where $\rho_r$ and $\rho_g$ are the mass densities of the rubber and glass particles, respectively. The volume fraction of the rubber particles is denoted with $\nu$.

\subsection{XRCT system and oedometer cell}
\label{sec:XRCT_system_oedometer_cell}
\begin{figure}[htb]
	\centering
	\includegraphics[width=1.0\textwidth]{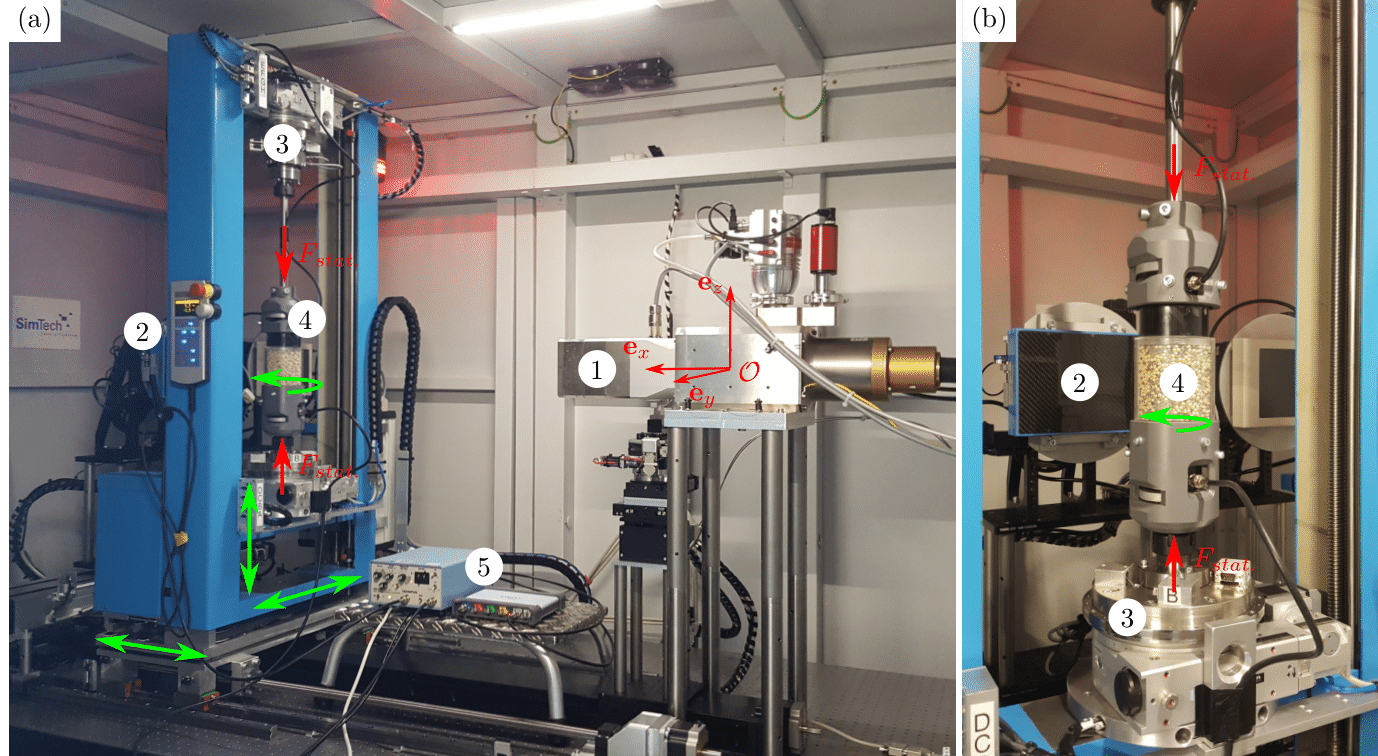}
	\caption{Experimental in situ setup: (1) X-ray source, (2) X-ray flat panel detector, (3) Universal testing machine with synchronized rotary tables, (4) Oedometer cell with integrated P-wave transducers containing monodisperse glass and rubber particles, (5) Square wave/pulser receiver unit and PC oscilloscope for ultrasonic measurements.}
	\label{fig:experimental_setup}
\end{figure}

The experimental study was performed in an open, modular, and flexible lab-based micro X-ray Computed Tomography (\textmu{XRCT}) system with an intrinsic large space allowing for the installation of a mechanical testing device required for in situ investigations. A detailed description of the system can be found in \cite{Ruf2020}. The system presented in \cite{Ruf2020} was extended by the integration of a load frame for XRCT investigations, cf. Figure~\ref{fig:experimental_setup}(a). 
For this, a Zwick 1445 10 kN Universal Testing Machine (UTM), refurbished from Doli Elektronik GmbH, Germany, and equipped with a modern EDC 222V controlling system, is employed. In the UTM, two rotatory tables with high loading capacities (XHuber 1-Circle Goniometer 411-X3W2) are integrated. The rotatory tables are pre-stressed with the controlled force $F_{stat.}$ applied to the sample during the image acquisition process. The X-ray source of the system is an open micro-focus tube with tungsten transmission target (FineTec FORE 180.01C TT) from FineTec FineFocus Technology GmbH, Germany. For the image acquisition, a Dexela detector 1512NDT with GOS-based DRZ Standard scintillator option from PerkinElmer, Inc., Waltham, MA, USA is applied.

To perform ultrasonic through transmission measurements in situ, a low X-ray absorbing oedometer cell was designed and manufactured in-house. Compared to a triaxial cell, only the vertical stress in the axial direction is controlled. In Figure~\ref{fig:experimental_setup} the oedometer cell installed in the XRCT-system and loaded by the employed universal testing device is presented. 
\begin{figure}[htb]
	\centering
	\includegraphics[width=0.6\textwidth]{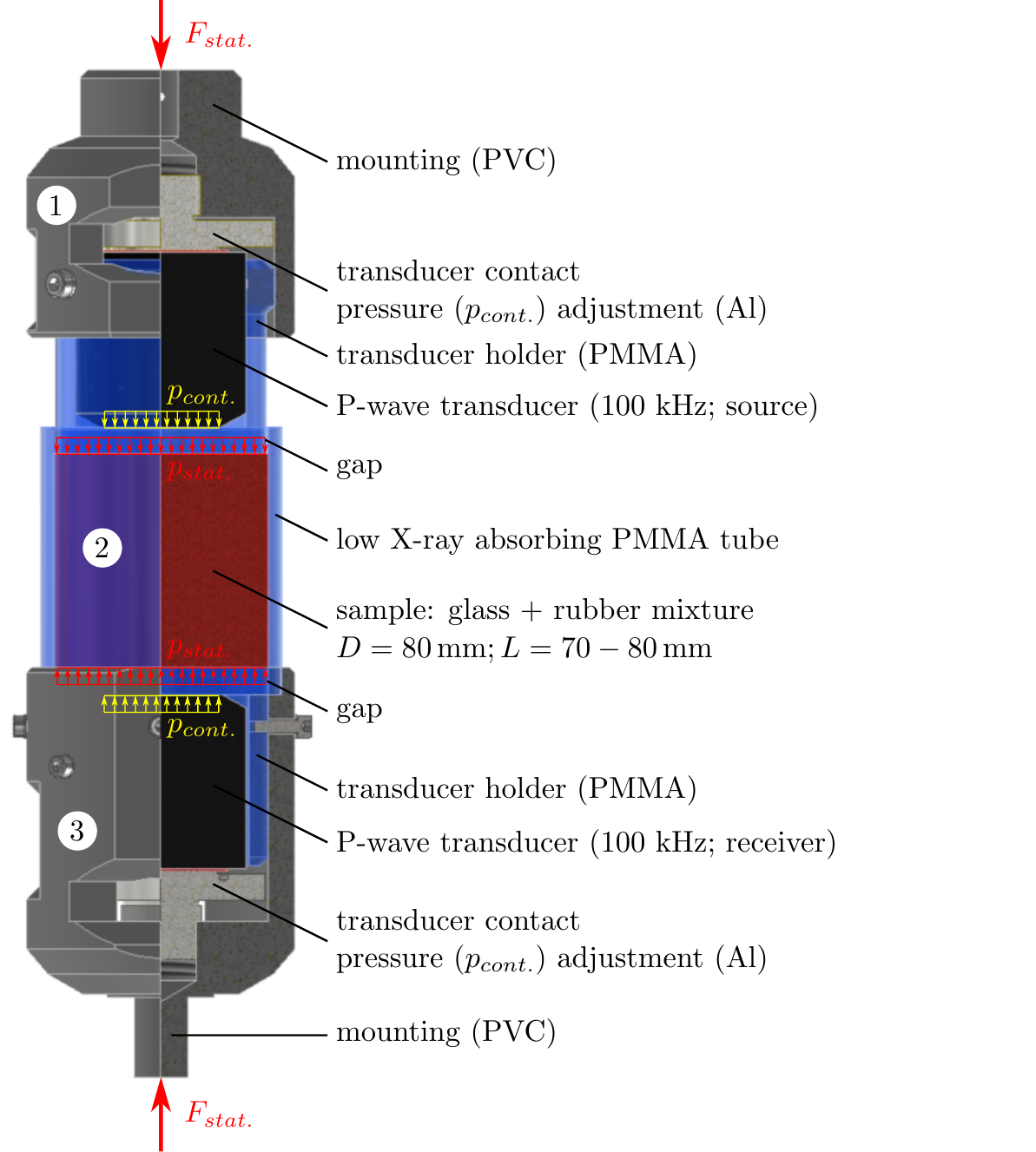}
	\caption{Three-quarter section view of the low X-ray absorbing oedometer cell with integrated ultrasonic transducers for in situ measurements.}
	\label{fig:experimental_setup_oedometer_cell}
\end{figure}
Further, the electrical devices for the ultrasonic testing corresponding to the schematic Figure~\ref{fig:wave_setup} can be observed. A detailed view of the oedometer cell design is provided in Figure~\ref{fig:experimental_setup_oedometer_cell}. The oedometer cell consists of three components. The top (1) and bottom (3) parts, both with integrated P-wave transducers, transfer the applied axial compression force $F_{stat.}$ to the sample and results in the axial stresses denoted as pressure $p_{stat.}$. As transducers, the pair of \mbox{100 kHz} P-wave broadband piezoelectric transducers (Olympus-Panametrics Videoscan V1011) is employed, cf. Subsection~\ref{sec:Stiffness_measurement}. For the coupling of the transducers to the Poly (methyl methacrylate) (PMMA) holders of the oedometer cell, an adequate couplant fluid is used. On both sides the wave travels through a \mbox{10\,mm} long PMMA distance, before and after the wave enters and leaves the sample. The contact pressure $p_{cont.}$ in between the transducers and the PMMA holders can be adjusted by an adjusting wheel. The sample is held in a rigid confining ring (2) made out of PMMA which prevents lateral displacement of the investigated mixture. Between the PMMA ring and the top and bottom part is a small gap in the radial direction to ensure that the emitted waves propagate through the sample. PMMA is used for the confining ring, as it has a very low attenuation coefficient. The inner diameter $D$ of the PMMA ring is $\mbox{80 mm}$. The sample height $H=L$ can be varied from \mbox{70 mm} to \mbox{80 mm}.

 Two capture the whole cell content (diameter 80 mm; height 80 mm), a geometric magnification $M_{geo.} = 1.36$ was set for all particulate systems. This leads to a field of view of 106.92 mm $\times$ 84.48 mm. Since we are not interested in small features (bead diameter 4 mm) all projection images were acquired in $2 \times 2$ detector binning mode. This means, that the detector's full resolution of \mbox{1944 $\times$ 1536} pixels with \mbox{74.8 {\textmu}}m pixel pitch and 14-bit pixel depth is reduced to \mbox{972 $\times$ 768} pixels with a pixel size of \mbox{149.6 {\textmu}}m. On one side, this significantly improves the signal-to-noise ratio (SNR), and on the other side, the final tomogram data set size is reduced by factor 8. Both simplify the subsequent image processing significantly without the loss of information required for the presented study. For all scans, an acceleration voltage of \mbox{110 kV} with an acceleration flux of \mbox{110 {\textmu}A} for the X-ray source were set. Combined with a detector exposure time of \mbox{1000 ms} and 1440 equidistant projection angles. Further, a detector bad pixel compensation as described in \cite{Ruf2020} was employed using two different projection positions for each projection angle. The final tomogram stacks have a size of 972 $\times$ 972 $\times$ 768 voxels with the uniform voxel edge length of 110 \textmu{m}. The 3d reconstruction of all scans was performed with the software Octopus Reconstruction (Version 8.9.4-64 bit) \cite{Vlassenbroeck2007} using the Filtered Back Projection (FBP) method \cite{Kak1987} in combination with the FDK reconstruction algorithm \cite{Feldkamp1984}. To account for common artifacts in XRCT imaging (ring artifacts and beam hardening), different types of filters were used. Based on the tomograms, segmentation is performed to distinguish between the rubber and glass beads and the remaining pore space. The segmentation workflow is described in detail in Section~\ref{section:image_analysis}.

\subsection{Image analysis}
\label{section:image_analysis}
\begin{figure}[hb]
	\centering
	\includegraphics[width=0.4\textwidth]{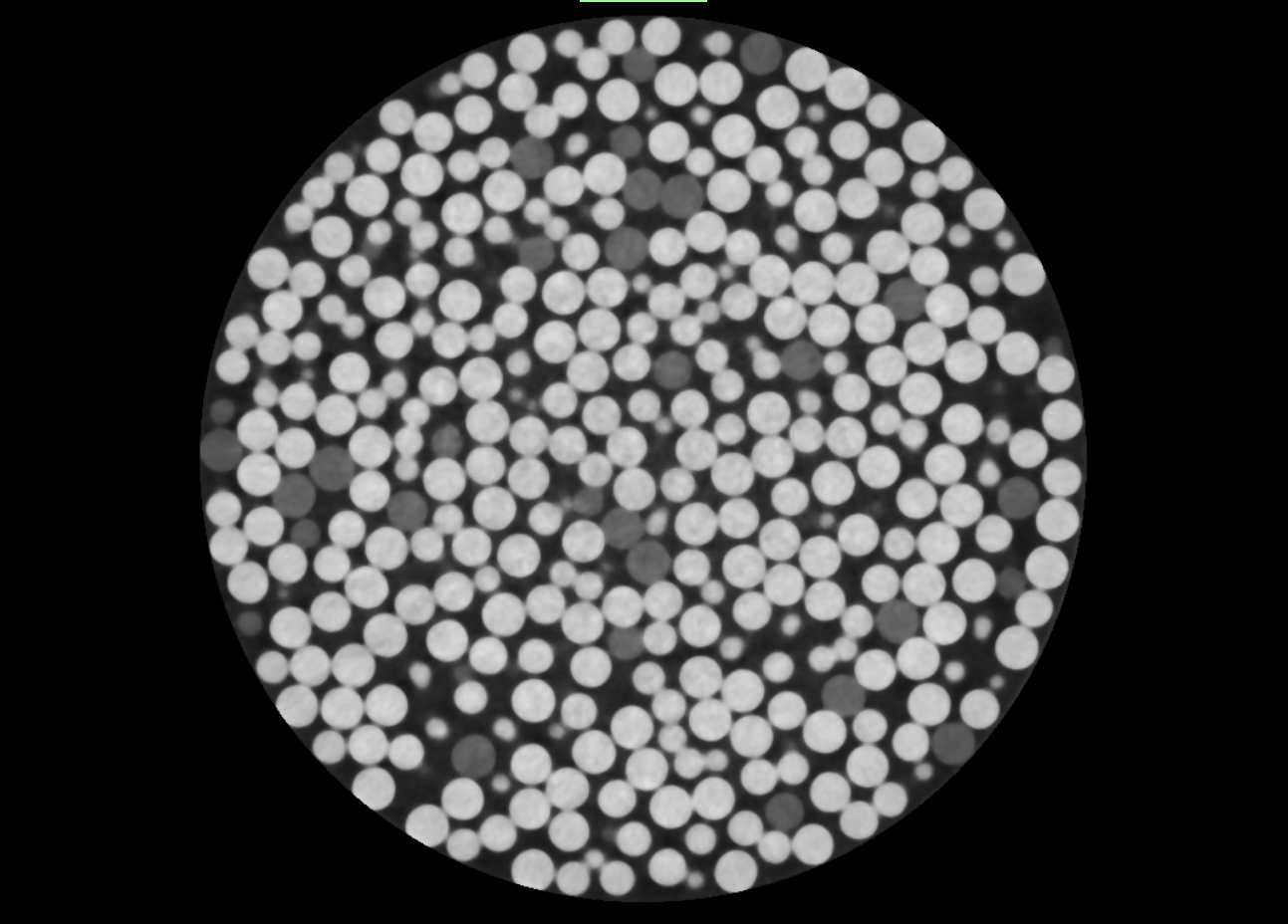}
	\includegraphics[width=0.4\textwidth]{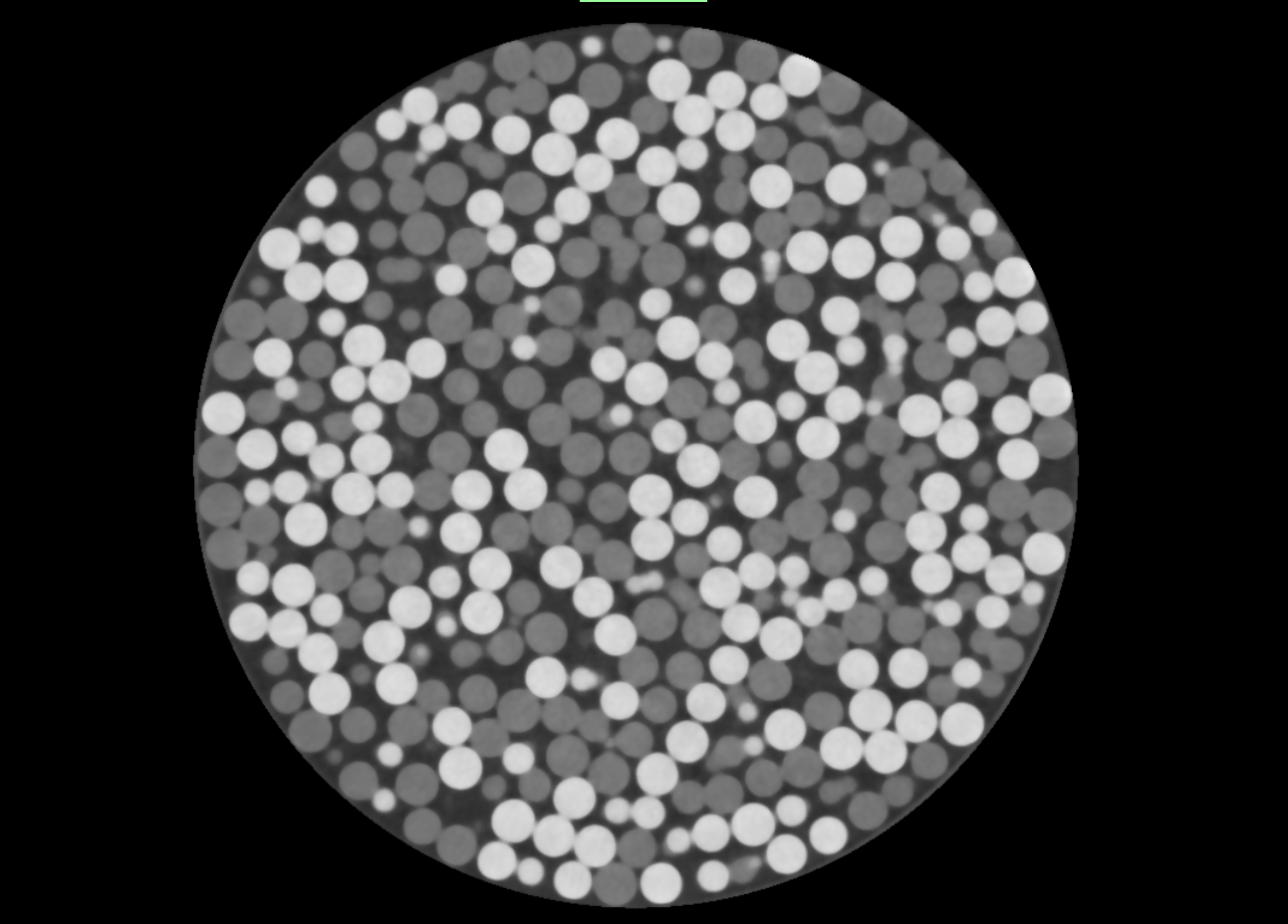}\\
	    \includegraphics[width=0.4\textwidth]{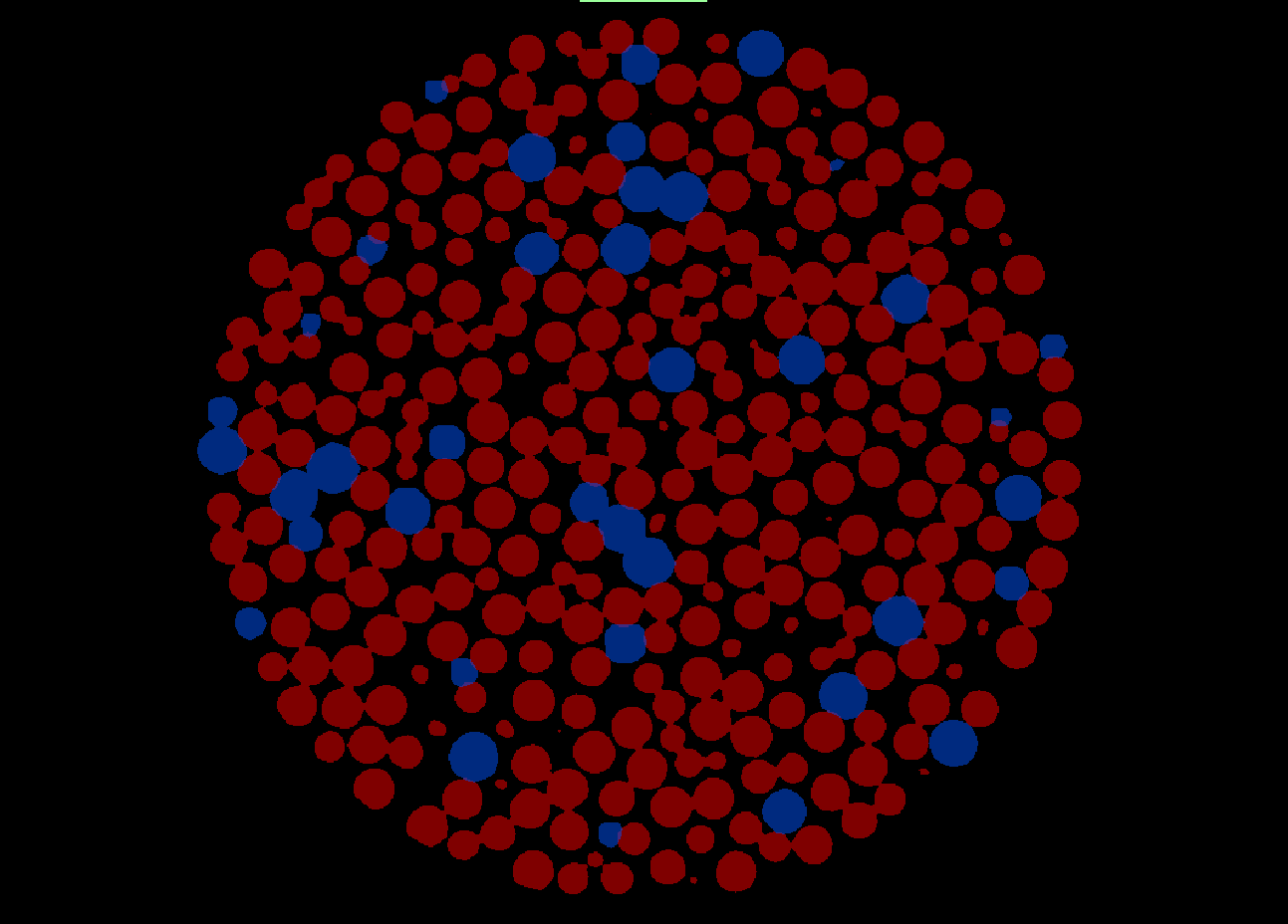}
	\includegraphics[width=0.4\textwidth]{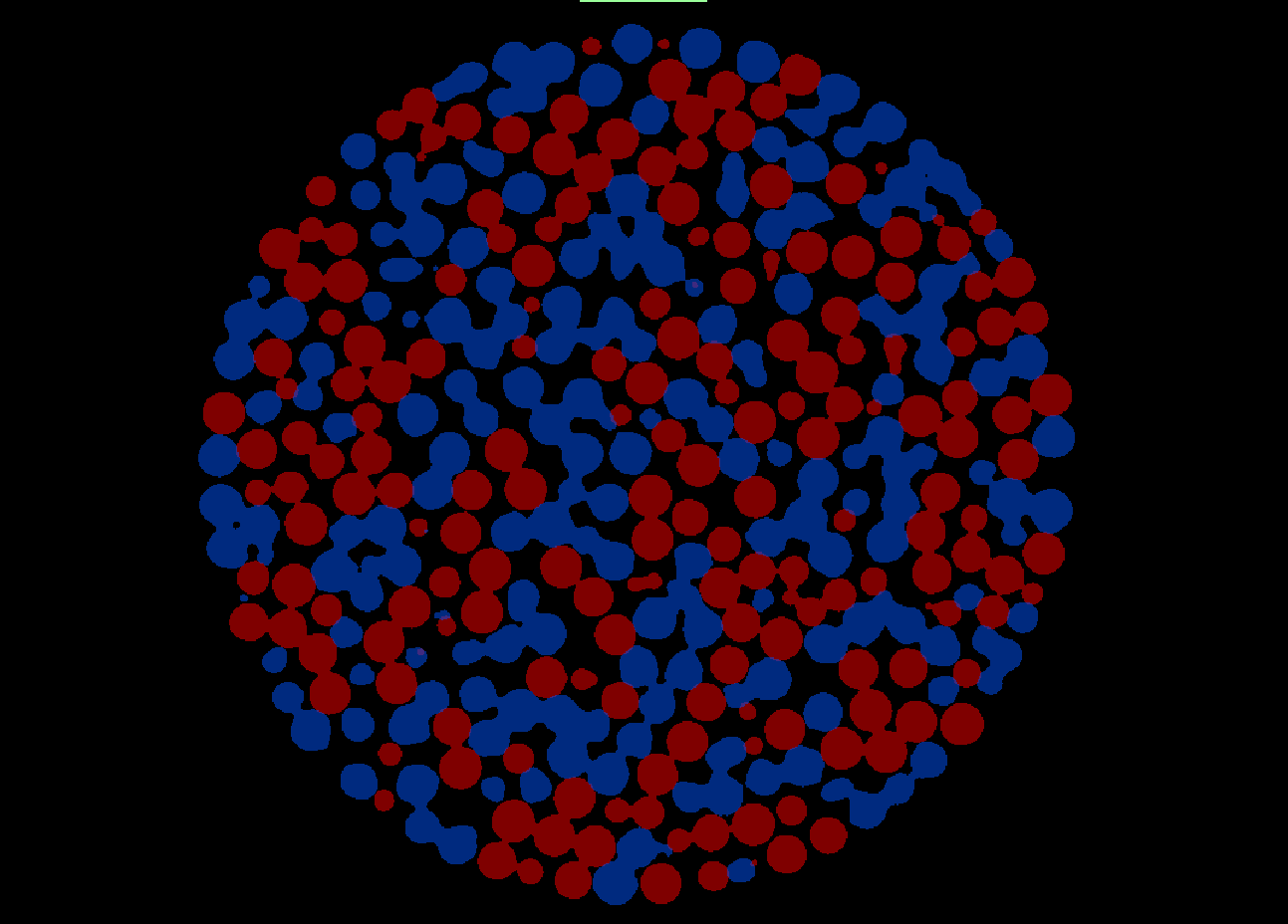}\\	
		\includegraphics[width=0.4\textwidth]{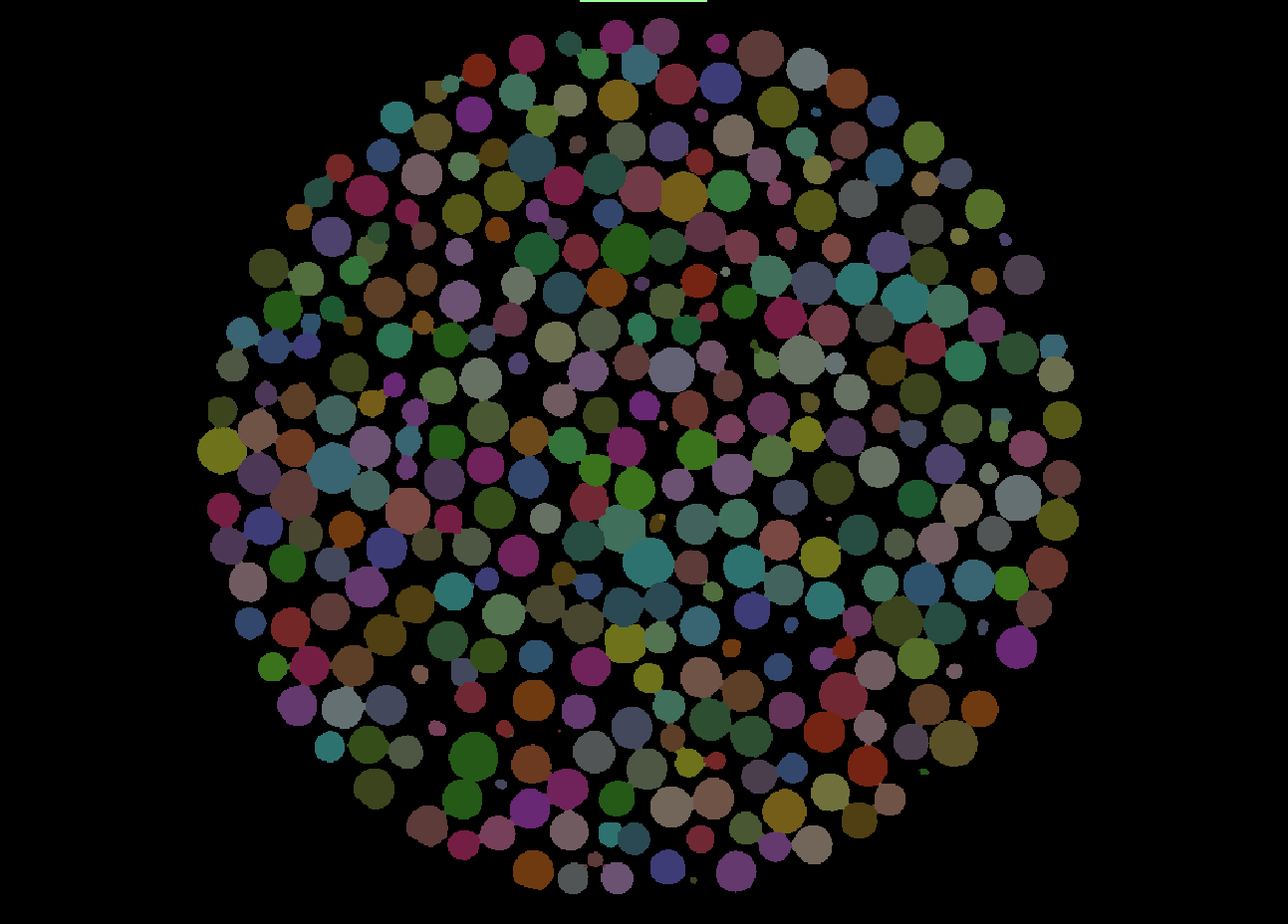}
	\includegraphics[width=0.4\textwidth]{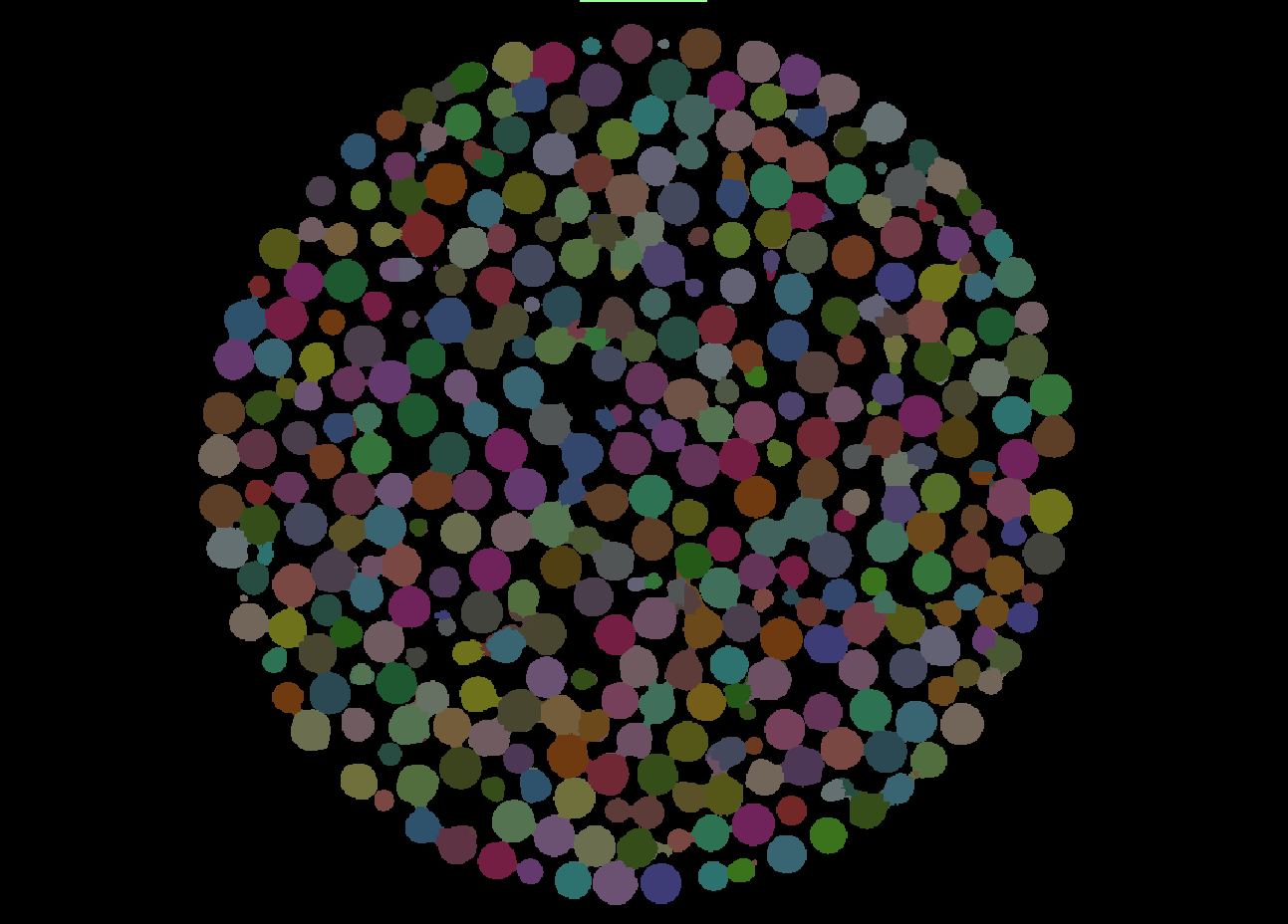}\\
	\caption{A top view of a xy-layer of particles for samples prepared with $\nu=0.1$ (left column) and 0.5 (right column). Top to bottom shows the steps taken from a reconstructed raw image (first row) to segmented (second row), and labeled (third row). The 3d raw data set for $\nu=0.5$, here presented with a single slice (``slice{\_}xy{\_}0322.tif''), can be found in \cite{Ruf2021}.}
	\label{fig:image_processing_workflow}
\end{figure}
\begin{figure}[hb]
	\centering
	\includegraphics[width=1.0\textwidth]{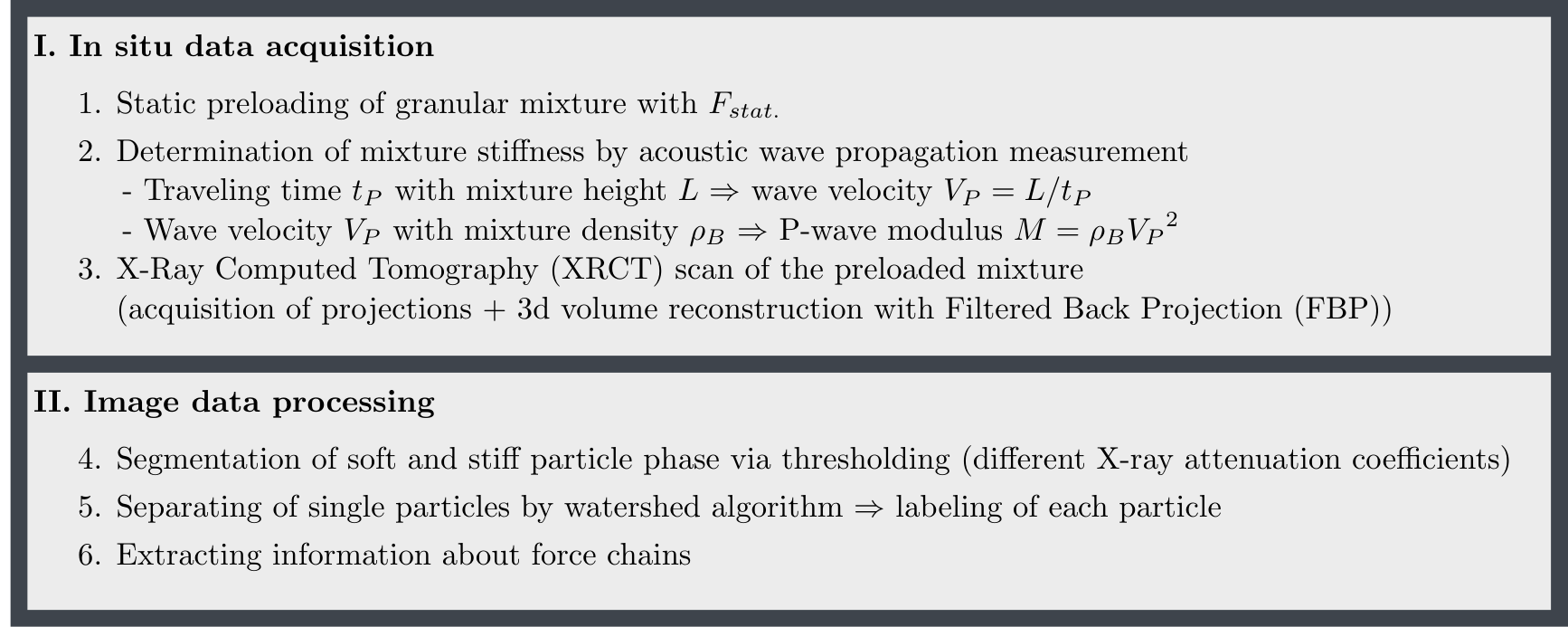}
	\caption{In situ workflow summary.}
	\label{fig:workflow_summary}
\end{figure}
Here, we present steps taken using the Dragonfly software version 2020.2 for Windows from Object Research Systems (ORS, Montreal, QC, Canada) \cite{Dragonfly}, for the segmentation and the subsequent quantification of mixture samples. 
An important step before interpreting the data is to identify individual particles. To increase the quality of images and reduce unwanted noise, one can apply different image filtering techniques before segmenting the phases, i.e.~separating different objects in an image. Here, a 3d-median smoothing function with a kernel size of seven was applied on image stacks to improve the resolution of pixels. Next, pixels of images were manually divided into three different regions of interest (air, rubber and glass) which are used as boundaries of objects. After segmenting samples into three separate regions (air, rubber, and glass), watershed transform technique is employed on defined markers of individual particles to label every particle in the separated regimes of rubber and glass. The watershed is a classical region-based algorithm that has its origins in mathematical morphology \cite {beucher1992watershed,digabel1978iterative,serra1986introduction,vincent1991watersheds} used for segmentation. Starting from user-defined markers, the watershed algorithm treats pixels values as a local topography (elevation). The algorithm floods basins from the markers until basins attributed to different markers meet on watershed lines. In many cases, markers are chosen as local minima of the image, from which basins are flooded.

Figure~\ref{fig:image_processing_workflow} shows a z-direction view of a layer of particles prepared at two rubber fractions, $\nu=0.1$ (left) and 0.5 (right), under \mbox{800 N} load in the oedometer cell. Light and dark gray pixels represent glass and rubber particles respectively in the first row. Second row shows the segmented pixels in the regions of interests (rubber (blue) and glass (red)) for which deep learning was employed. Thanks to the watershed transform technique, particles of segmented regions are labeled individually (third row).

\subsection{Investigated particulate samples and measurement workflow}
All investigated samples are made of monodisperse glass and rubber beads with \mbox{4 mm} diameter. Further information on the material characteristics for both, glass and rubber beads, can be found in \cite{taghizadeh2021elastic}. Particles are poured carefully into the oedometer cell described in Section~\ref{sec:XRCT_system_oedometer_cell}. Samples are prepared at different rubber volume fractions, $\nu=v_r/v_c$, from $\nu=0$ to 1 in 0.1 increments, where $v_r$ is the volume possessed by the rubber particles and $v_c$ is the total volume of particles in oedometer cell. All mixtures are tested in the custom-made oedometer cell with the sample height and diameter equal to \mbox{80 mm}, cf. Figure~\ref{fig:experimental_setup} and Figure~\ref{fig:experimental_setup_oedometer_cell} ex-situ without imaging. The prepared samples are uniaxially compressed in the axial direction via the top piston of the oedometer cell in subsequent force increments from $F = \mbox{200 N}$ to \mbox{1000 N}. At each instant step, the system was relaxed for some time at the defined force to overcome the creep behavior of soft particles. Then, at each intermediate load step, a high voltage burst signal (P-wave) is excited from the top cap (the sound source) transducer and the bottom transducer collects the signal. To remove the influence of network configurations and user errors on outputs, each experiment was repeated at least 3 times by mounting and dismounting particles into the cell. For the calculation of the system stiffness besides the wave propagation time~$t_P$ the exact height~$L$ of the compressed packing is required. This is determined based on the loading protocol of the UTM. 
Since about \mbox{2 h} are needed for one XRCT scan, only a subset of the possible combinations of the parameter space (rubber volume fraction $\nu$ and preload $F_{stat.}$ were scanned in situ. XRCT imaging was performed for a preload $F_{stat.} = \mbox{400 N}$ and $F_{stat.} = \mbox{800 N}$ in combination with rubber volume fractions $\nu = \{0.1, 0.2, 0.3, 0.4, 0.5, 0.6 \}$. Otherwise, the procedure is identical to the previously performed ex-situ measurements. During the image acquisition, it is switched from force-control to displacement-control to avoid slight movements of the particles due to relaxation processes. The selection is based on previous knowledge from comparable experiments in a triaxial cell without XRCT imaging \cite{taghizadeh2021elastic} as well as the before performed ex-situ measurements in the oedometer cell. A summary of the applied in situ workflow is given in Figure~\ref{fig:workflow_summary}.

\section{Results - insight into granular media}
\label{sec:results}
Measurement of ultrasound velocity $V_P$ provides complementary information about material properties and combined use of velocity and topological structure in seismic analysis provides greater insight into the granular packings. Here, we first report the results on the bulk stiffness of granular mixtures with diverse rubber content obtained by the experimental tests. Particular attention is devoted to the dependence of sound velocity on the applied load and soft-stiff composition of samples, since this is an important controllable experimental parameter \cite{kim2008sand,lee2007behavior,taghizadeh2021elastic}. After that, X-ray images taken during wave propagation are analysed to obtain micro-insights of packings.

\subsection{Material characterization by ultrasonic measurements}
\begin{figure}[htb!]
\subfigure(a){\includegraphics[scale=0.65,angle=0]{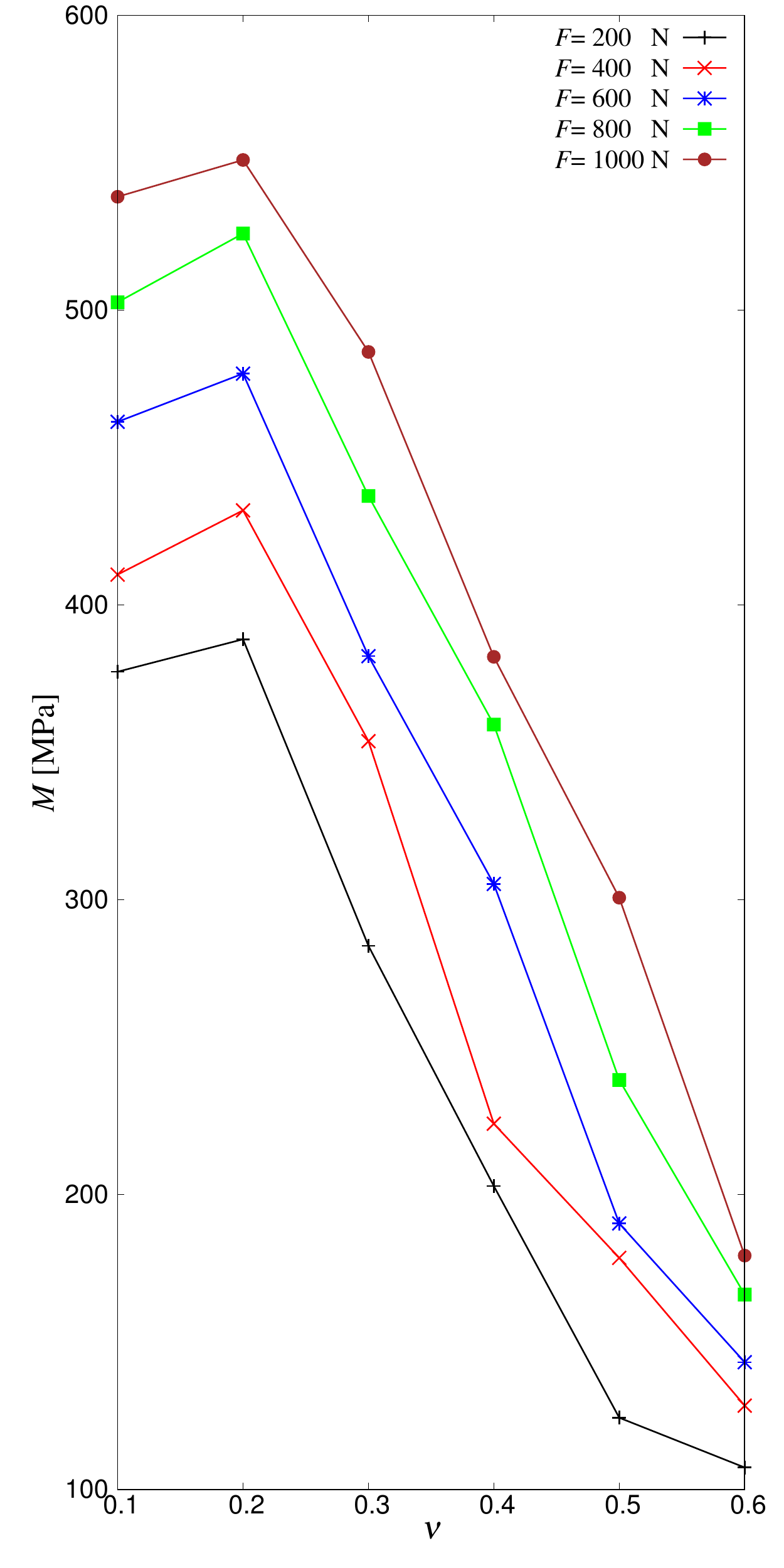}}
\subfigure(b){\includegraphics[scale=0.65,angle=0]{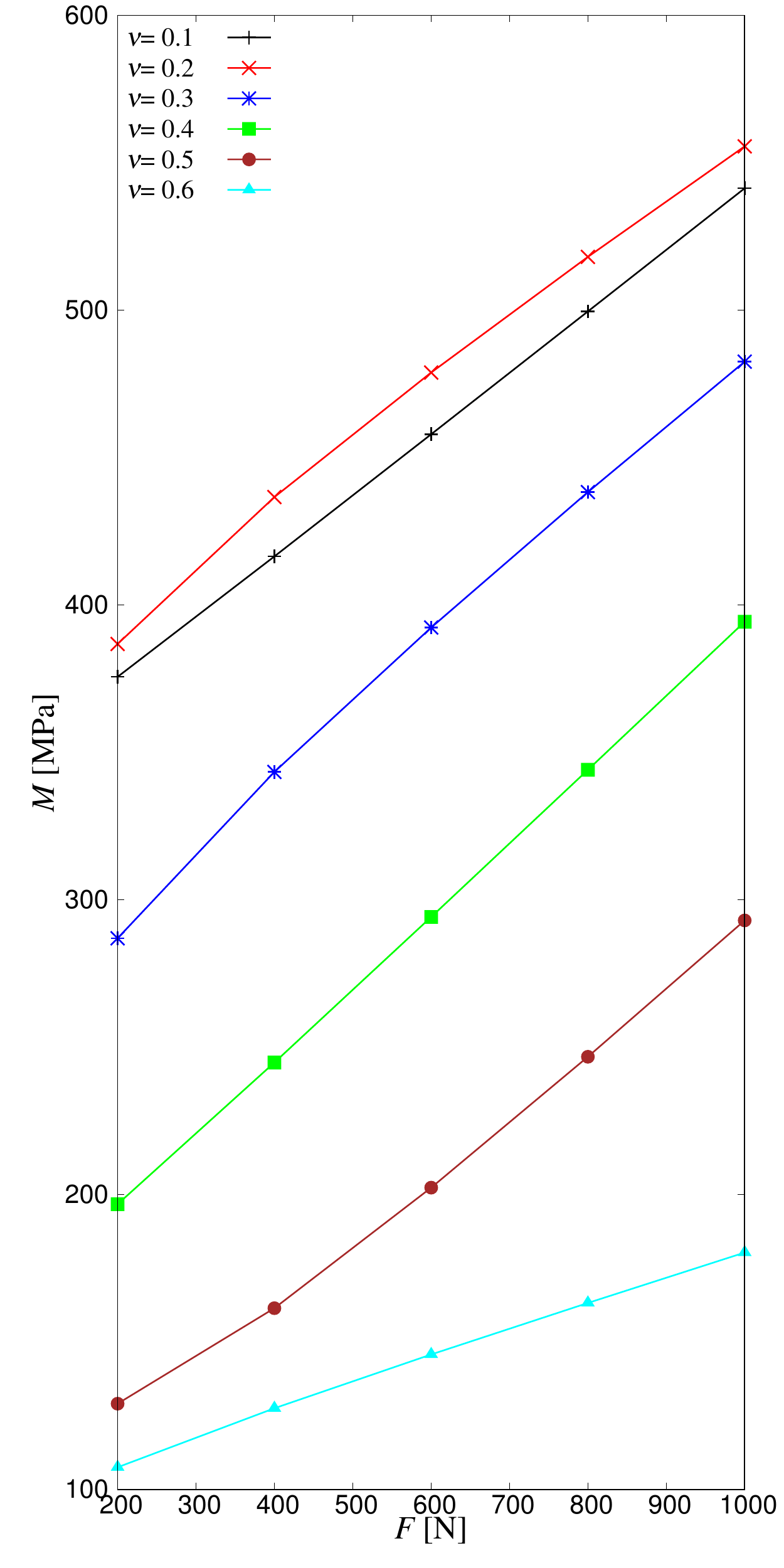}}\\
\caption{(a) In situ measured P-wave $M$ modulus against rubber fraction $\nu$. (b) P-wave modulus against uniaxial applied load from \mbox{$F=200$\,N} to \mbox{1000\,N}. The underlying measurement data for $\nu = 0.5$ is provided in \cite{Ruf2021}.}
\label{fig:results_plots}      
\end{figure}

Figure~\ref{fig:results_plots} shows the compressional modulus determined for samples prepared with rubber fractions from \mbox{$\nu =$ 0.1} to 0.6 at five different load steps acquired in situ. The underlying data of the mixture with $\nu = 0.5$ together with the recorded XRCT data set is for illustrative purposes available in \cite{Ruf2021}. $M$ modulus shows high stiffness for rubber content up to $\nu=0.3$ where the bulk behavior of samples is highly controlled by the stiff phase. Previous experimental studies under triaxial stress conditions showed already that a small amount of soft particles enhances the effective stiffness of the medium \cite{taghizadeh2021elastic}. Thus, the effective stiffness of bidisperse granular mixtures consisting of stiff and soft particles does not follow a simple mixture rule. The highest modulus is observed at $\nu \approx$ 0.2. Thus, granular mixtures can be manipulated to obtain aggregates with even higher stiffness, but lighter and more dissipative thanks to rubber, when appropriate external conditions are matched (in this case the pressure) \cite{kim2008sand}. Increasing the amount of rubber particles, here 0.3 < $\nu$ < 0.6, reduces the effective stiffness where a phase transition from stiff to soft phase occurs.

\subsection{Image characterization}
\begin{figure}[hp]
	\centering
	\subfigure(a){\includegraphics[width=0.43\textwidth]{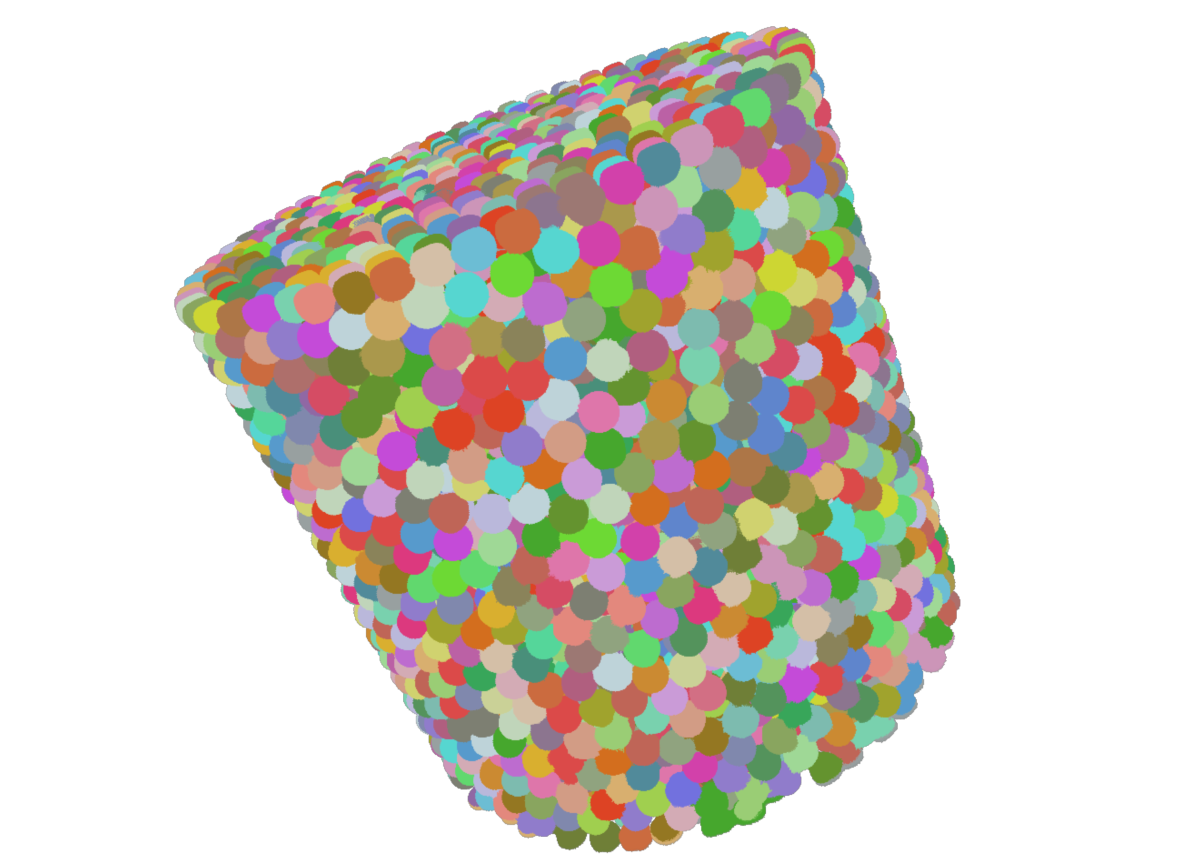}}
	\subfigure(b){\includegraphics[width=0.45\textwidth]{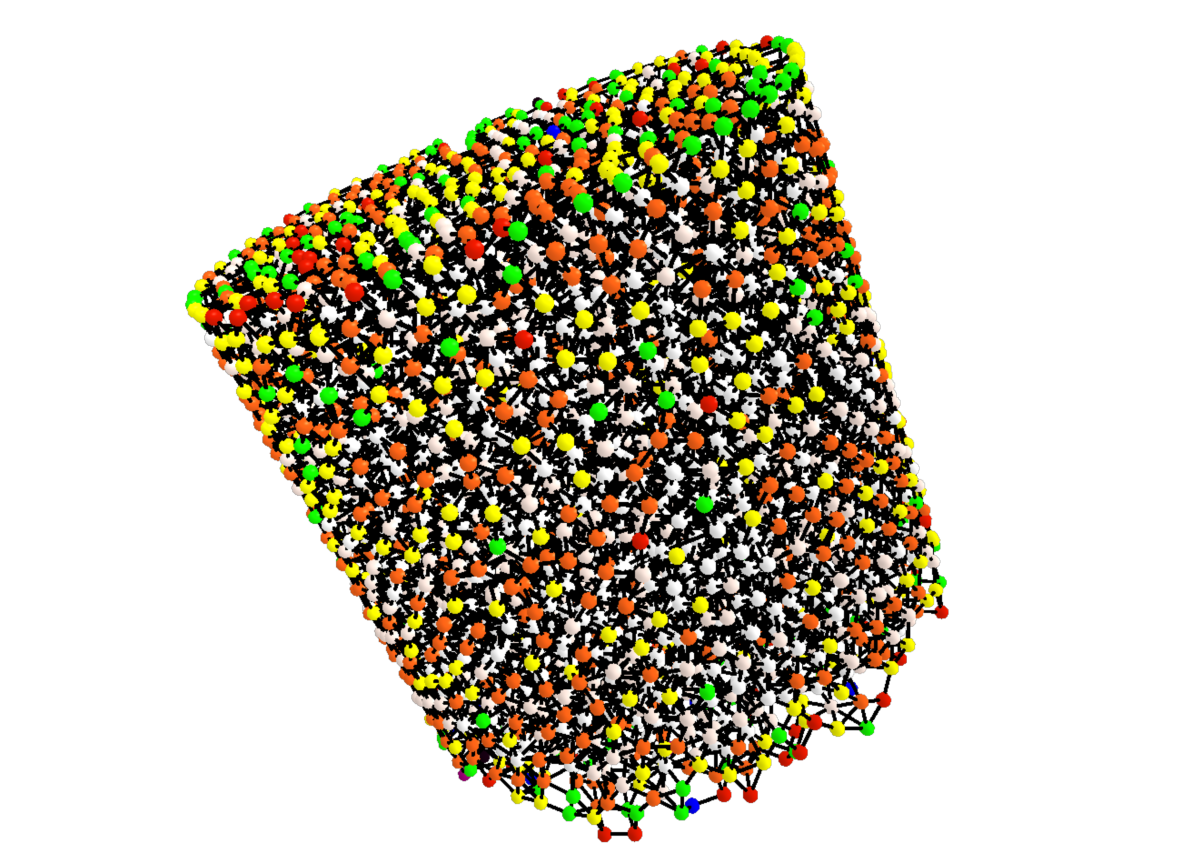}}
	\caption{(a) 3d view of labeled particles for a sample prepared with $\nu=0.5$. (b) The network of contacts of the sample in which different particles colors represent a number of contact each particle carries. Particles carrying zero, one, two, three, four, five, six, and above six number of contacts are colored by black, magenta, blue, red, green, yellow, orange, and white, respectively.
	}
	\label{fig:labeled3d}
	\centering
	\subfigure(a){\includegraphics[width=0.43\textwidth]{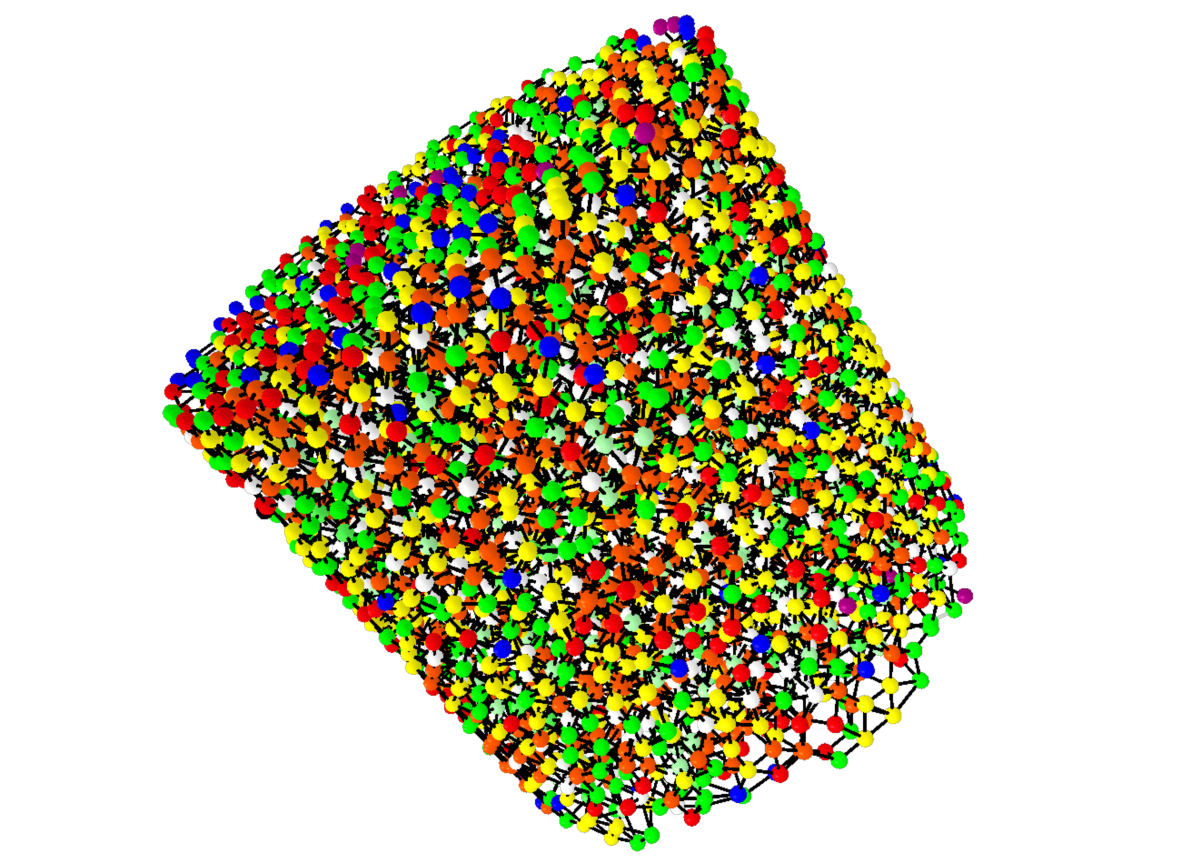}}
	\subfigure(b){\includegraphics[width=0.45\textwidth]{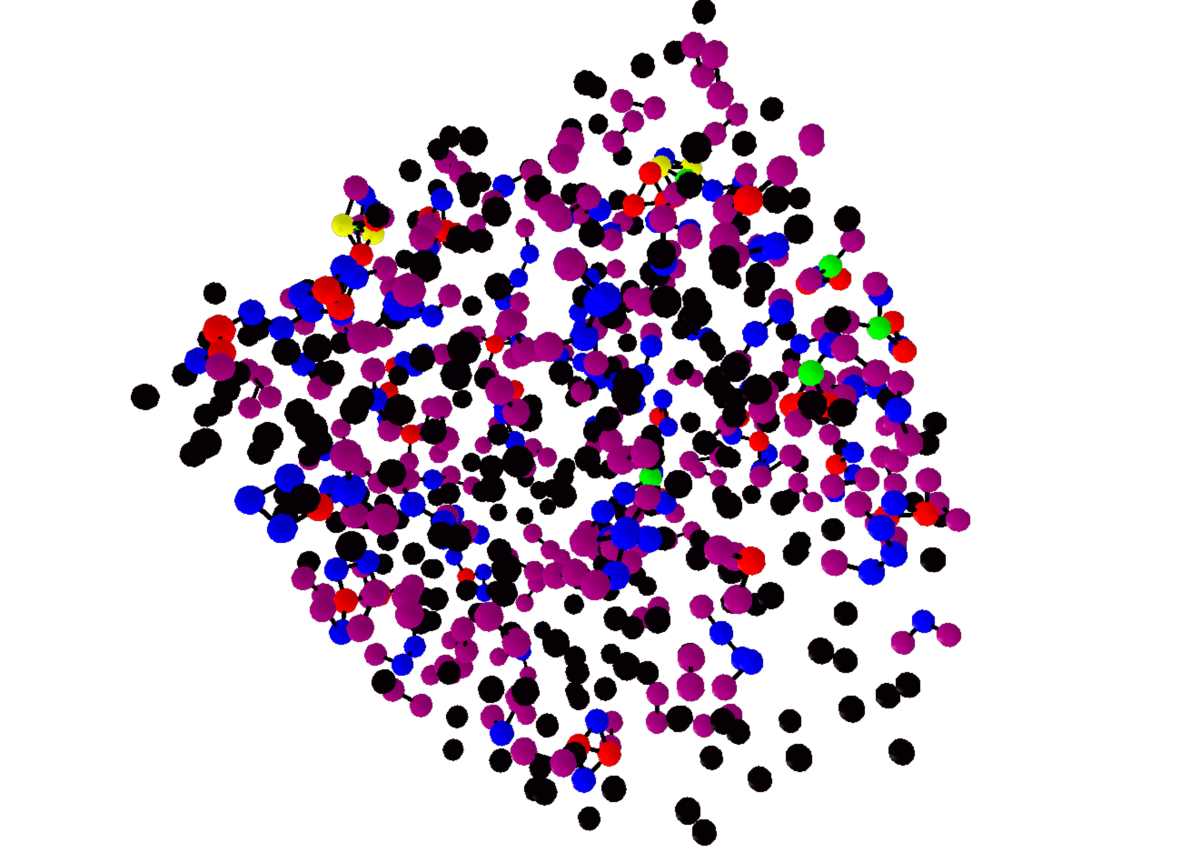}}
	\caption{3d view of (a) glass and (b) rubber particles network for a sample prepared with rubber fraction $\nu=0.1$, respectively. Particles center are marked with different colors based on their number of contacts and connected with black lines. Particles carrying zero, one, two, three, four, five, six, and above six number of contacts are colored by black, magenta, blue, red, green, yellow, orange, and white, respectively.}
	\label{fig:3dnetwork10}
	\centering
	\subfigure(a){\includegraphics[width=0.43\textwidth]{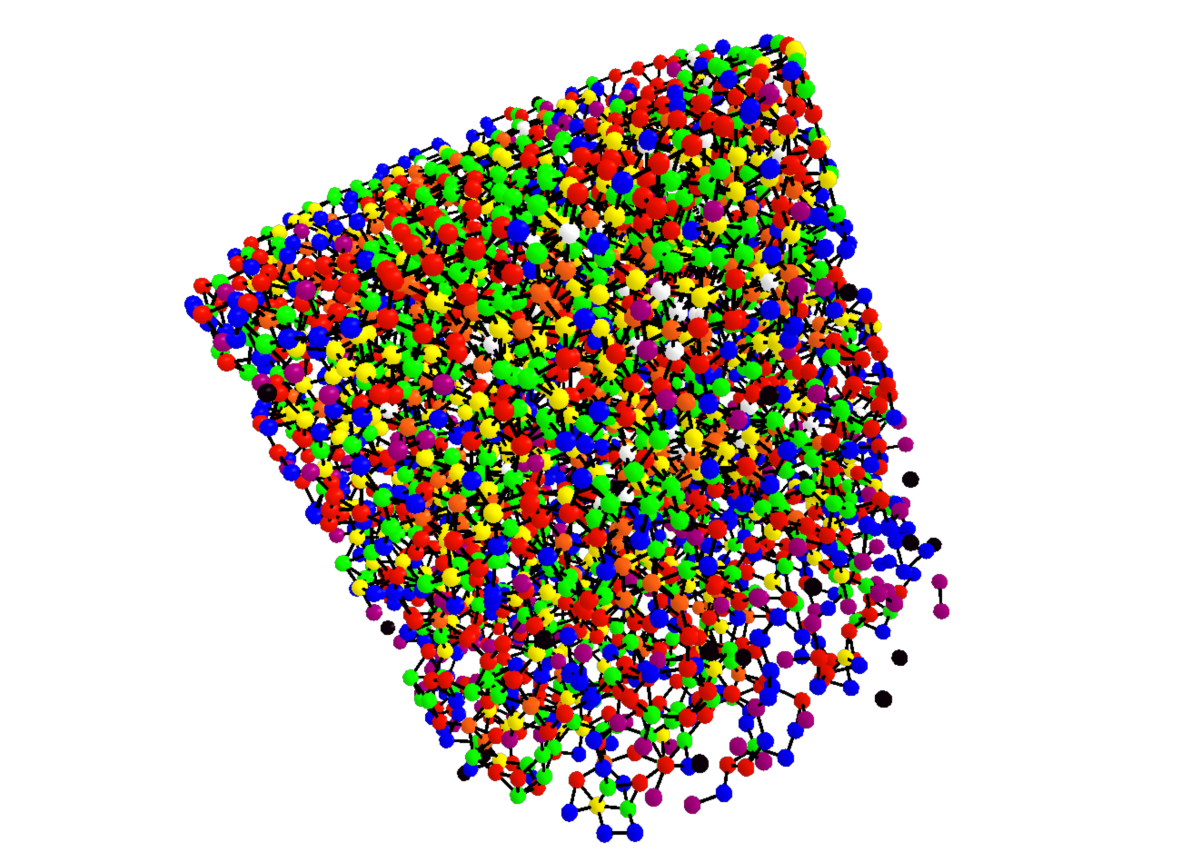}}
	\subfigure(b){\includegraphics[width=0.45\textwidth]{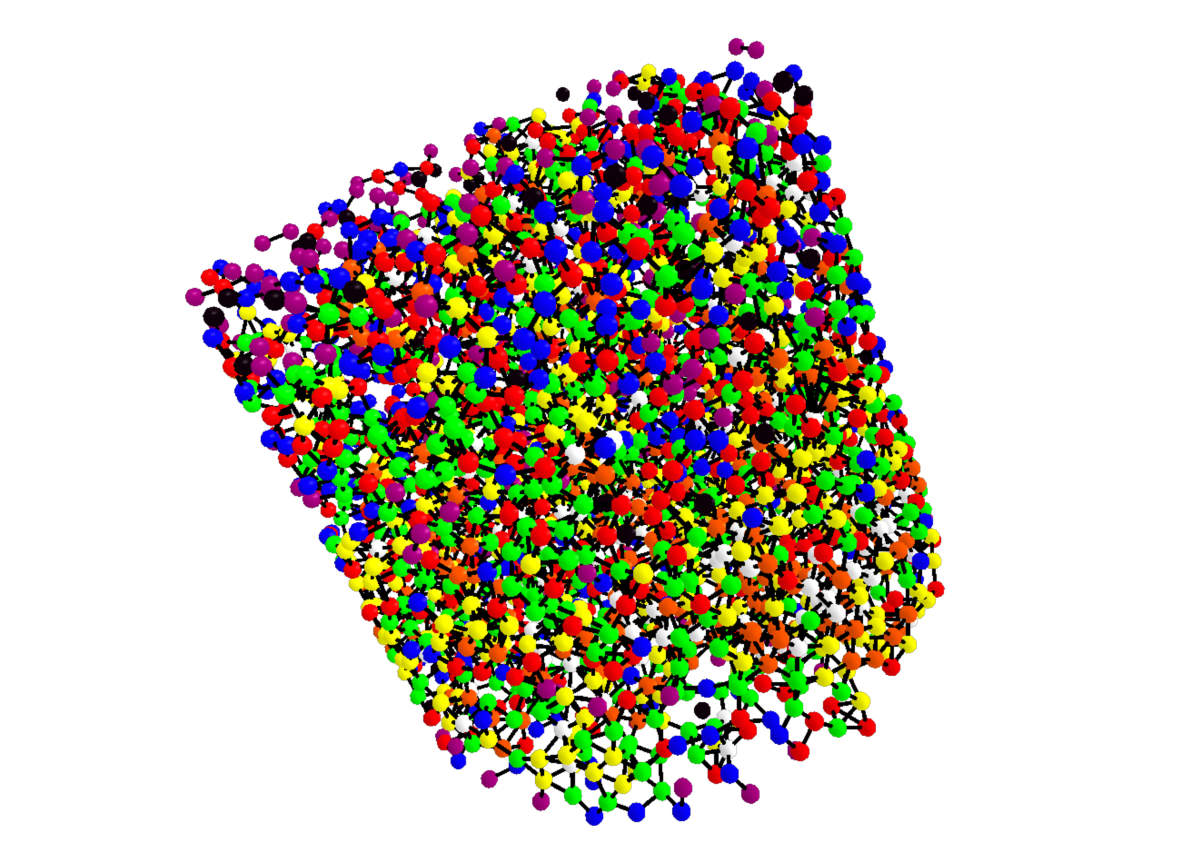}}
	\caption{3d view of (a) glass and (b) rubber particles network for a sample prepared with rubber fraction $\nu=0.5$, respectively. Particles center are marked with different colors based on their number of contacts and connected with black lines. Particles carrying zero, one, two, three, four, five, six, and above six number of contacts are colored by black, magenta, blue, red, green, yellow, orange, and white, respectively.}
	\label{fig:3dnetwork50}
\end{figure}
Recent progresses in the application of $\mu$XRCT imaging in the field of geomechanics have allowed all the individual particles of packings in a test sample to be seen and identified uniquely in 3d. Combining such imaging capabilities with experiments carried out ``in situ'' within an imaging setup (shown in Figure~\ref{fig:experimental_setup}) has led to the possibility of directly observing the topology of packings. 

Granular matter has a heterogeneous nature, this is expressed as force chains through the medium. These force chains spread the forces through the contact with another particle \cite{taghizadeh2021stochastic,taghizadeh2021energy}. This causes that neighboring particles could have forces of different magnitudes and non-isotropic distributions of stress throughout the medium \cite{bassett2012influence}. This heterogeneity is manifest of the fact that granular matter exhibits a strong configuration and history dependence. One of the interesting outcomes of XRCT imaging of particulate systems is its potentiality in depicting a particles contact network (e.g.~force chains) which reveals micro-structural information. A 3d view of labeled particles is shown in Figure~\ref{fig:labeled3d}(a) for a sample prepared with $\nu=0.5$; and the contact network of the sample is illustrated in Figure~\ref{fig:labeled3d}(b). Center of particles, marked with different colors, are connected with black lines. The color code of the number of contacts follows as black (zero), magenta (one), blue (two), red (three), green (four), yellow (five), orange (six), and white (above six). 

Next, to enhance micro-structural investigation, glass and rubber particles are separated as a result of labelling every particles for samples with volume fraction of $\nu=0.1$ (Figure~\ref{fig:3dnetwork10}) and 0.5 (Figure~\ref{fig:3dnetwork50}). Looking at glass (Figure~\ref{fig:3dnetwork10}(a)) and rubber (Figure~\ref{fig:3dnetwork10}(b)) networks of $\nu=0.1$, it is not surprising to see the scatter of rubbers among glass beads. These rubbers which do not carry number of contacts above three are called rattlers. Rattlers do not transfer force between each others as their number of contact are not sufficient to form a chain of particles; whereas, moving to a sample with higher rubber fraction $\nu=0.5$ (Figure~\ref{fig:3dnetwork50}), we do not see rattles neither in the glass and nor in the rubber networks. This observation explains why $M$ modulus remains fairly constant for samples with low rubber volume fraction ($\nu \leq 0.3$) in Figure~\ref{fig:results_plots} since their bulk behavior is controlled by stiff particles. But in case of intermediate regime, $\nu$ between 0.4 and 0.6, forces are distributed among glass and rubber chains which leads to a smaller $M$ modulus.

\section{Discussion}
\label{sec:discussion}
As is almost always the case, no experimental investigation is perfect. The problems that arise must be critically questioned with regard to their influence on the scientific question. The motivation for the presented study was given by investigations of the same kinds of mixtures in a conventional triaxial cell with well-defined boundary conditions \cite{taghizadeh2021elastic}. Since it is quite more complicated to design and build such a cell that can be used in an XRCT device, we went for an oedometer cell. As the results in Figure~\ref{fig:results_plots} show, we can observe the same phenomenon. As a consequence, it is an admissible simplification for the problem under investigation.

One common significant uncertainty and difficulty is associated with determining the exact wave propagation time by ultrasonic measurements required to calculate the sample stiffness. Suggested criteria and recommendations vary depending on the installation, application, and input signal. The most common methodology is to interpret the received signal in the time domain. It is typical to consider the first peak at the receiver transducer as arrival time and the required time difference minus the needed time in the other parts as the travel time of the signal within the packings \cite{o2015analysis} which was consistently done within this study. However, in particular, for high rubber packings, choosing a peak is not easy since the wave contains high-frequency tones raised from the cluster of rubber particles. Thus, a low-pass filter was applied to remove unwanted noises. For low rubber volume fractions, this is nevertheless a difficult undertaking and not completely objective. Although the same experimental protocol and measurement were applied, a user error is inevitable. To account for this, the before performed ex-situ measurements were repeated three times and are consistent with the in situ acquired data shown in Figure~\ref{fig:results_plots}.

To avoid movements of the particles during the image acquisition period taking up to \mbox{2 h}, it was switched from force-control to displacement-control. This was done to avoid slight movements of the particles due to creeping processes caused by the (viscoelastic) rubber particles. As a result, a corresponding relaxation can be observed in the measured load, which, however, is less critical with regard to the movement of individual beads. This can be confirmed by the final quality of the 3d images as well as by comparing the in situ ultrasonic measurement results with the before performed ex-situ measurements. Consequently, the relatively long XRCT scan time has no influence on the imaging. In the shown concept, the static preload $F_{stat.}$ is applied using a load frame, here a modified traditional universal testing machine. Consequently, two rotary tables must be used to rotate the pre-stressed sample. While this concept generally offers extremely great flexibility (uniaxial tension/compression and torsion) compared to specially designed load cells with integrated actuators, cf. \cite{Buffiere2010,Singh2014} and the literature cited therein, it also presents some difficulties. The major problem is the accurate alignment of the two rotatory tables in the micrometer range. Each eccentricity error leads either to an undefined stress state and/or to bad image quality due to potential movement errors in soft samples since the error is multiplied with the set geometric magnification ($M_{geo.}$). However, because an extremely low geometric magnification is applied here (\mbox{$M_{geo.} = 1.36$}) and, in addition, the individual particles are comparatively large with a diameter of \mbox{4 mm}, this influence is to be classified as minor on the underlying scientific question. Although it is slightly present in the samples with a very low rubber fraction. To get completely rid of this error source the alignment of the rotary tables should be further improved which is 
technically challenging or alternatively, a load cell with integrated actuator could be employed. Latter leads however to a significantly more complicated and expensive cell design. That is the reason why this approach was intentionally not used. Common artifacts in XRCT imaging (e.g. noise, ring artifacts, system alignment errors, and beam hardening) could be reduced to a minimum by the combined usage of hardware precautions and software-based corrections. In general, it is advantageous to use relatively large particles as in our approach, as this significantly reduces many potential sources of error. For instance, alignment errors are less significant and here a $2 \times 2$ detector binning was possible which significantly reduces the intrinsic noise in the data sets. This simplifies the subsequent image processing due to a significant better signal-to-noise ratio. 

Segmentation of XRCT data is doubtlessly the most critical step before quantitative analyses. Due to the good quality of the raw data sets (justifying the scanning time of \mbox{2 h}), and the fact that only three high-contrast phases (air, glass, and rubber) had to be segmented, this could be accomplished with a straight forward traditional workflow. The separation of the single particles was performed using the well-known watershed algorithm. Since the exact composition of each mixture is known (total particle volume and number of particles), the segmentation and subsequent separation of the single particles could be validated quite easily on a global level. The exact deformation of the individual particles is not of interest for the presented study. Consequently, the image processing and the quantification can be judged as very reliable. 

It can be concluded that the presented in situ setup in combination with the applied workflow is adequate to get a more comprehensive understanding of particulate systems as presented in this contribution. Further, it demonstrates that an investigation of such kinds of questions are accessible with a laboratory-based XRCT system if some boundary conditions are fulfilled, e.g. an intrinsic large space within the system.

\section{Summary}
\label{sec:summary}
Understanding the response of granular-based systems in applications requires a detailed grasp of the connection between the basic ingredients (particles) and the macroscale properties of the systems considered. These are complex systems and an understanding of the overall behavior cannot be gained by studying individual particles. While significant progress has been made during the last decades on understanding relevant physical mechanisms, there are still many open questions, starting from the physics of particle interactions to general features of multiscale models that will bridge the different spatial and temporal scales of interest. The purpose of this research was to describe an in situ approach for the combination of traditional experimental techniques in combination with 3d imaging in order to explore a micro-macro relation of granular assemblies.

The given contribution started initially with a short overview about attenuation-based XRCT imaging provided in Section~\ref{sec:image_characterization}. Based on this, a detailed explanation of how X-ray computed tomography and wave propagation measuring technique can be combined was demonstrated in Section~\ref{sec:In-situ_experimental_testing}. For this, a low X-ray absorbing oedometer cell with integrated P-wave transducers was designed and built. The cell was integrated into a modular {\textmu}XRCT-system which was extended by a load frame using a refurbished universal testing machine. This approach enables applying in a flexible manner different kinds of load cases needed for advanced mechanical in situ investigations. A smart and elegant approach to segment different material phases of the 3d tomograms was explained step-by-step. The sample material chosen for the study is a composition of monodisperse granular particles, glass, and rubber beads; not only because of the ubiquity of granular materials but also because it is a paradigm for complex disordered media and the unique characteristics exhibited by various physical phenomena associated with mechanical waves in it (dispersion, scattering attenuation, intrinsic attenuation, diffusion, weak localization, energy transfer across different frequencies, etc.).

In Section~\ref{sec:results}, the effect of rubber volume content variation of the studied glass-rubber mixtures on the compressive elastic modulus (P-wave modulus) by means of wave propagation at different uniaxial preload levels was shown. It was demonstrated that the modulus of the mixtures can be increased by adding low portions of soft particles since it is dominated by stiff particles. Adding more soft particles into assemblies led to a transition from stiff to soft dominated regime where bulk behavior of samples is mainly controlled by soft phase. The macro behavior of such granular samples can hardly be described without micro information. To gain more insights, XRCT imaging was performed in situ for two of the five examined preloads. Analyzing images helped to understand the transition mechanism from stiff to soft controlled regimes by separating glass and rubber networks. It was found that due to the isolation of rubber particles, rattlers, the samples were dominated by glass grains. Whereas, a sample with \mbox{$50\,\%$} of rubber and glass has distributed number of contacts, fairly, among themselves.

The combination of material and image characterization experimental techniques is a complex procedure that was explained carefully in this article. However, it could be shown what incredible potential it has. Especially since many problems from the field of mechanics are accessible in laboratory-based XRCT systems. Numerical simulations, based on the Discrete Element Method (DEM), have revealed the outmost role of the microstructure in characterizing the elastic behavior of granular soils. To further advance our understanding, one could simulate particles networks obtained by image processing using DEM.

\section*{Acknowledgments}
The authors acknowledge the scientific discussion with Stefan Luding. Also, we thank Ralf Plonus for his technical support in modification of devices. M.R. and H.S. acknowledge funding from the German Science Foundation (DFG) through Project No. STE 969/13-1. K.T. and H.S. acknowledge funding by the German Science Foundation (DFG) through the project STE-969/16-1 within the SPP 1897 ``Calm, Smooth and Smart''. H.S. thanks the DFG for supporting this work under Grant No. SFB 1313 (Project No. 327154368).

\subsection*{Conflict of interest}
The authors declare no potential conflict of interests.

\section*{Supporting information}
For demonstrative purposes, the {\textmu}XRCT images (reconstructed data sets, projection data sets, and metadata) of the investigated glass-rubber mixture with a rubber volume content \mbox{$\nu = 0.5$} as well as the corresponding ultrasonic measurement data are openly available in the Data Repository of the University of Stuttgart (DaRUS) at \url{https://doi.org/10.18419/darus-2208}, \cite{Ruf2021}.

\bibliographystyle{abbrvnat}
\bibliography{.//literature.bib}

\end{document}